\documentclass[fleqn,usenatbib]{mnras}

\usepackage{newtxtext,newtxmath}

\usepackage[T1]{fontenc}

\usepackage{newtxtext,newtxmath}
\usepackage{multirow}
\usepackage[flushleft]{threeparttable}
\usepackage{booktabs}

\usepackage{xspace}
\def\thisgrb{GRB\,241105A\xspace}
\def\gbmTninty{{$69.0$\xspace}}

\DeclareRobustCommand{\VAN}[3]{#2}
\let\VANthebibliography\thebibliography
\def\thebibliography{\DeclareRobustCommand{\VAN}[3]{##3}\VANthebibliography}

\usepackage{newtxtext,newtxmath}
\usepackage{fontawesome5}
\usepackage{mathtools}
\usepackage{float}
\usepackage{graphicx}
\usepackage{xtab,booktabs}
\usepackage{longtable}
\usepackage{lipsum}
\usepackage{natbib}
\usepackage[dvipsnames]{xcolor}

\newcommand{\tninty}{{$T_{\rm 90}$}\xspace}
\newcommand{\swift}{{\it Swift}\xspace}
\newcommand{\fermi}{{\it Fermi}\xspace}
\newcommand{\svom}{{\it SVOM}\xspace}
\newcommand{\gbm}{{\it Fermi}/GBM\xspace}
\newcommand{\grm}{{\it SVOM}/GRM\xspace}
\newcommand{\bat}{{\it Swift}/BAT\xspace}
\newcommand{\xrt}{{\it Swift}/XRT\xspace}
\newcommand{\uvot}{{\it Swift}/UVOT\xspace}
\newcommand{\keV}{{\rm keV}\xspace}

\newcommand{\feii}{\mbox{Fe\,{\sc ii}}}

\newcommand{\siii}{\mbox{Si\,{\sc ii}}}
\newcommand{\sii}{\mbox{S\,{\sc ii}}}

\newcommand{\alii}{\mbox{Al\,{\sc ii}}}
\newcommand{\aliii}{\mbox{Al\,{\sc iii}}}

\newcommand{\cii}{\mbox{C\,{\sc ii}}}
\newcommand{\civ}{\mbox{C\,{\sc iv}}}
\newcommand{\siiv}{\mbox{Si\,{\sc iv}}}

\newcommand{\oi}{\mbox{O\,{\sc i}}}

\newcommand{\nv}{\mbox{N\,{\sc v}}}

\title[\thisgrb]{\thisgrb: A test case for GRB classification and rapid r-process nucleosynthesis channels}
\author[Dimple et al.]{ 
Dimple$^{1,2}$\thanks{E-mail: d.dimple@bham.ac.uk},
B.~P.~Gompertz$^{1,2}$,
A.~J.~Levan$^{3}$,
D.~B.~Malesani$^{3}$,
T.~Laskar$^{9,3}$,
S.~Bala$^{4}$,
A.~A.~Chrimes$^{5,3}$,
\newauthor
K.~Heintz$^{6}$,
L.~Izzo$^{7}$,
G.~P.~Lamb$^{8}$,
D.~O'Neill$^{1,2}$,
J.~T.~Palmerio$^{10}$,
A.~Saccardi$^{10}$,
G.~E.~Anderson$^{11}$,
\newauthor
C.~De~Barra$^{12}$,
Y.~Huang$^{13}$,
A.~Kumar$^{14}$,
H.~Li$^{15}$,
S.~McBreen$^{12}$,
O.~Mukherjee$^{4}$,
S.~R.~Oates$^{16}$,
U.~Pathak$^{17}$,
\newauthor
Y.~Qiu$^{15}$,
O.~J.~Roberts$^{4}$,
R.~Sonawane$^{18,65}$,
P.~Veres$^{19}$,
K.~Ackley$^{20}$,
X.~Han$^{15}$,
Y.~Julakanti$^{20}$,
J.~Wang$^{15}$,
\newauthor
P.~D'Avanzo$^{21}$,
A.~Martin-Carrillo$^{12}$,
M.~E.~Ravasio$^{3,21}$,
A.~Rossi$^{22}$,
N.~R.~Tanvir$^{23}$,
J. ~P.~ Anderson$^{58,63}$,
\newauthor
M.~Arabsalmani$^{24,25}$,
S.~Belkin$^{32}$,
R.~P.~Breton$^{26}$,
R.~Brivio$^{21}$,
E.~Burns$^{62}$,
J.~Casares$^{27,28}$,
S.~Campana$^{21}$,
\newauthor
S.~I.~Chastain$^{29}$,
V.~D'Elia$^{30}$,
V.~S.~Dhillon$^{31,27}$,
M.~J.~Dyer$^{31}$,
J.~P.~U.~Fynbo$^{6}$,
D.~K.~Galloway$^{32}$,
A.~Gulati$^{33,34,35}$,
\newauthor
B.~Godson$^{20}$,
A.~J.~Goodwin$^{11}$,
M.~Gromadzki$^{60}$,
D.~H.~Hartmann$^{36}$,
P.~Jakobsson$^{37}$,
T.~L.~Killestein$^{20}$,
\newauthor
R.~Kotak$^{38}$,
J.~K.~Leung$^{39,40,41}$,
J.~D.~Lyman$^{20}$,
A.~Melandri$^{42}$,
S.~Mattila$^{38,43}$,
S.~McGee$^{1,2}$,
C.~Morley$^{11}$,
\newauthor
T.~Mukherjee$^{44,45}$,
T. ~E.~Müller-Bravo$^{55,56}$,
K.~Noysena$^{46}$,
L.~K.~Nuttall$^{47}$,
P.~O'Brien$^{23}$,
M.~De~Pasquale$^{48}$,
\newauthor
G.~Pignata$^{59}$,
D.~Pollacco$^{20}$,
G.~Pugliese$^{49}$,
G.~Ramsay$^{50}$,
A.~Sahu$^{20}$,
R.~Salvaterra$^{51}$,
P.~Schady$^{52}$,
B.~Schneider$^{53}$,
\newauthor
D.~Steeghs$^{20}$,
R.~L.~C.~Starling$^{23}$,
K.~Tsalapatas$^{57}$,
K.~Ulaczyk$^{20}$,
A.~J.~van~der~Horst$^{54}$,
C.~Wang$^{13,64}$,
\newauthor
K.~ Wiersema$^{61}$,
I.~Worssam$^{1,2}$,
M.~E.~Wortley$^{1,2}$,
S.~Xiong$^{13}$,
T.~Zafar$^{44}$}

\date{Accepted XXX. Received YYY; in original form ZZZ}

\pubyear{2025}

\begin{document}
\label{firstpage}
\pagerange{\pageref{firstpage}--\pageref{lastpage}}
\maketitle

\begin{abstract}
Gamma-ray bursts (GRBs) offer a powerful window to probe the progenitor systems responsible for the formation of heavy elements through the rapid neutron capture (\textit{r}-) process, thanks to their exceptional luminosity, which allows them to be observed across vast cosmic distances.  \thisgrb, observed at a redshift of $z = 2.681$, features a short initial spike ($\sim$1.5\,s) and a prolonged weak emission lasting about 64\,s, positioning it as a candidate for a compact binary merger and potentially marking it as the most distant merger-driven GRB observed to date. 
However, the emerging ambiguity in GRB classification necessitates further investigation into the burst’s true nature. Prompt emission analyses, such as hardness ratio, spectral lag, and minimum variability timescales, yield mixed classifications, while machine learning-based clustering places GRB 241105A near both long-duration mergers and collapsar GRBs. We conducted observations using the James Webb Space Telescope (JWST) to search for a potential supernova counterpart. Although no conclusive evidence was found for a supernova, the host galaxy's properties derived from the JWST observations suggest active star formation with low metallicity, and a sub-kpc offset of the afterglow from the host, which appears broadly consistent with a collapsar origin. Nevertheless, a compact binary merger origin cannot be ruled out, as the burst may plausibly arise from a fast progenitor channel. This would have important implications for heavy element enrichment in the early Universe. 

\end{abstract}




\section{Introduction}
Gamma-ray bursts (GRBs) are the most powerful explosions in the universe, releasing an enormous amount of energy in a short period. Historically, GRBs have been divided into two observational classes. Those with durations of up to two seconds (termed `short' GRBs) are thought to be powered by neutron star mergers, while those with longer durations (`long' GRBs) come from the collapse of very massive stars \citep[][see also \citealt{Zhang2004,Zhang2006,Zhang2009,Zhang2014a}]{Kouveliotou93}. However, the limitations of this framework have been recognized over time \citep{Fynbo06,Gehrels06,Zhang2009,Dimple_2022}. Recent advances have shown that this duration-based division does not cleanly separate mergers from collapsars. In particular, the discovery of kilonovae accompanying GRBs 211211A \citep{Mei22,Rastinejad22,Troja2022,Yang2022} and 230307A \citep{Gillanders23,Levan24,Yang_2024}, both with durations of $\sim 30$ s, has changed the canonical classification scheme. Such long-lasting merger-powered GRBs are incompatible with the short accretion timescale expected under the standard assumption that the merging neutron star binary promptly collapses to a black hole, and may indicate the need for additional physics like magnetic field effects \citep{Proga06,Gottlieb23}, a longer-lasting central engine \citep{Metzger11,Bucciantini12,Rowlinson13,Gompertz13,Gompertz14}, or the merger of a neutron star - black hole binary system \citep[e.g.][]{Desai19, Gompertz20, Dimple23}.

\begin{figure*}
    \centering
    \includegraphics[width=0.33\linewidth]{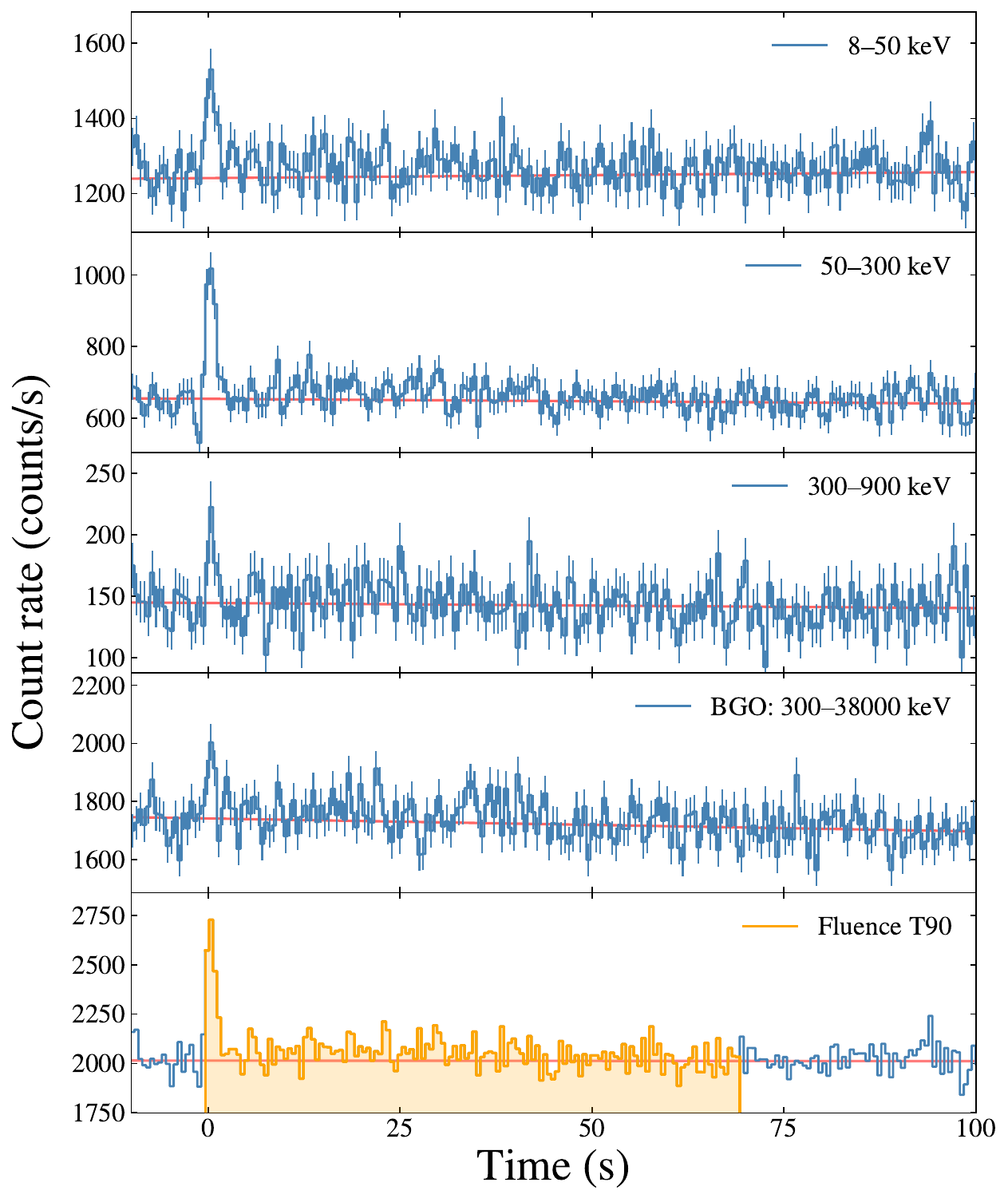}
    \includegraphics[width=0.33\linewidth]{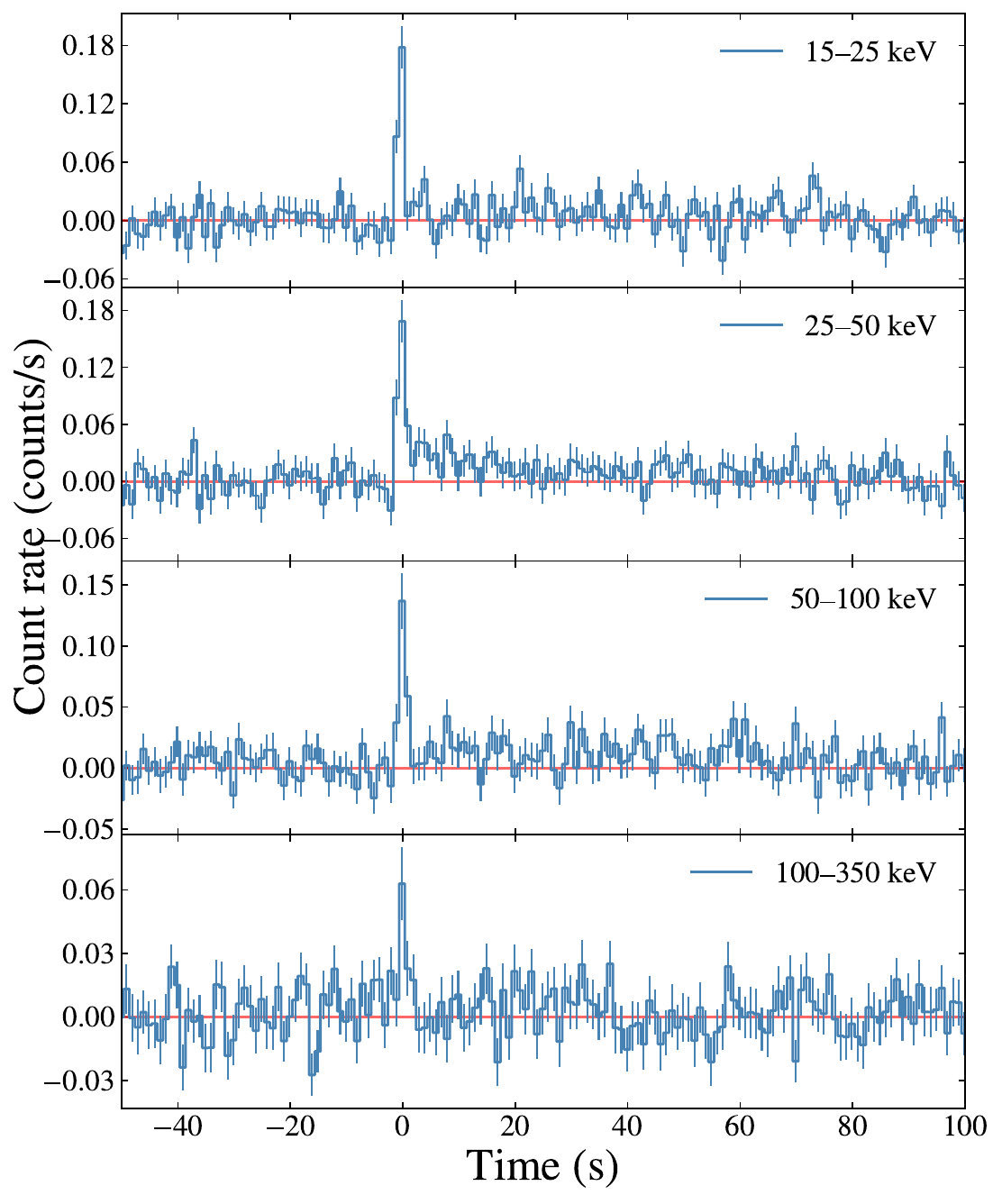}
    \includegraphics[width=0.33\linewidth]{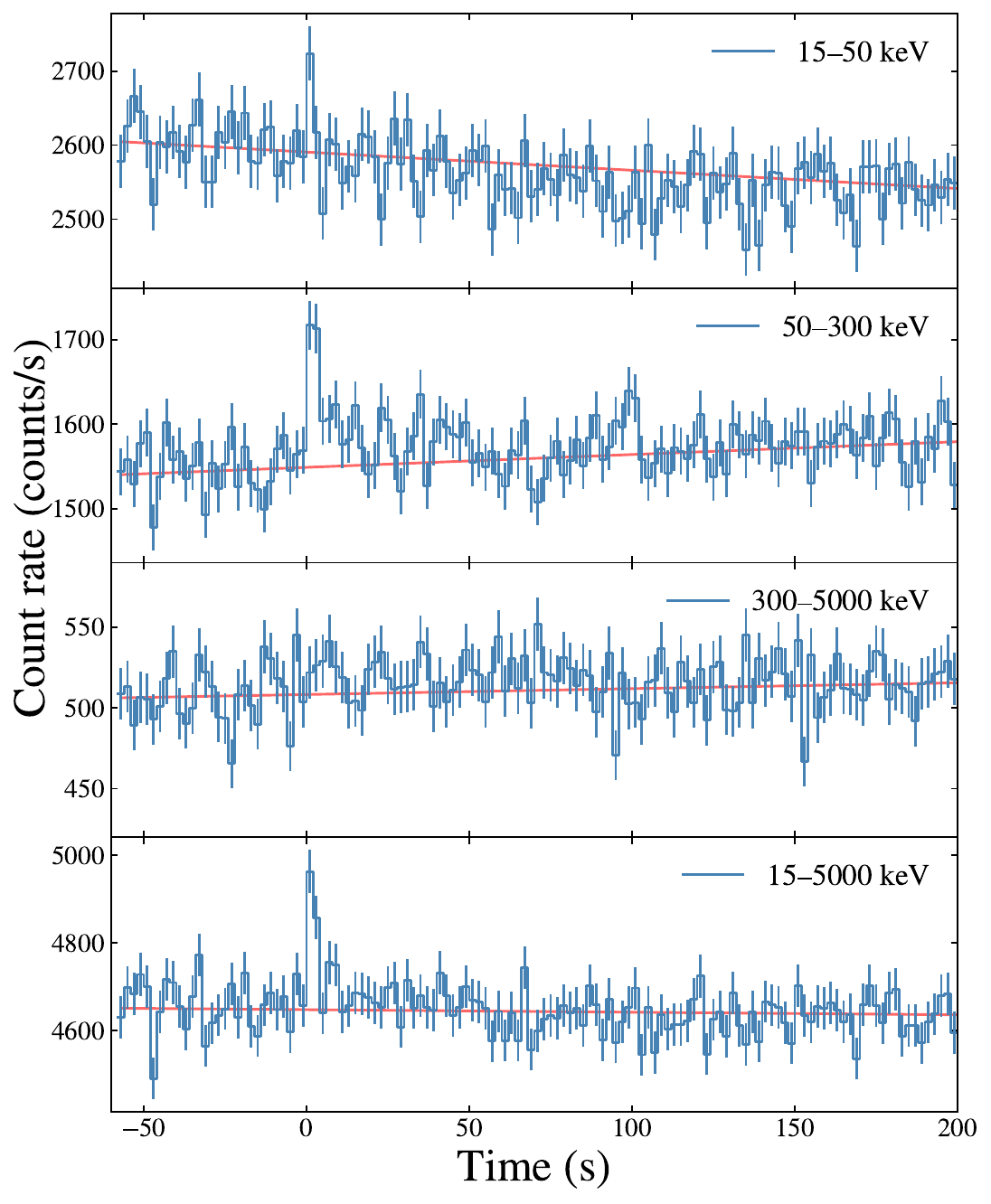}
    
    \caption{Light curves of GRB 241105A as detected by \gbm (left), \bat (middle), and \grm (right) in different energy bands. The red lines are the fitted background. The bottom panel in the GBM plot (left) shows \tninty of \gbmTninty ~s overlaid on the combined NaI light curve in the 8-900 keV range.}
    \label{fig:lc_241105A}
\end{figure*}
The current classification uncertainty has intensified interest in methods to reliably separate GRBs powered by collapsing stars from those powered by merging compact objects based on the prompt gamma-ray emission. Potential diagnostic properties include apparently distinct correlations between the isotropic equivalent energy emitted in gamma-rays ($E_{\rm iso}$) and the peak spectral energy \citep[$E_p$; the `Amati relation;'][]{Amati02,Amati06,Minaev20}; non-zero lags in the arrival times of soft gamma-ray photons relative to hard ones in collapsar GRBs \citep{Norris02}; and disparate minimum variability timescales (MVTs) between the two engine types \citep{MacLachlan13} which may probe the size of the emission region. Lately, efforts have turned to classifications based on machine learning (ML) techniques \citep{Chattopadhyay_2017, Acuner_2018, Jespersen20, Steinhardt2023, Salmon22, Dimple23, Mehta_2024, Dimple24, Zhu_2024}. The application of unsupervised ML algorithms on prompt emission light curves observed by the Burst Alert Telescope \citep[BAT;][]{Bat_barthelmy2005} on the \emph{Neil Gehrels Swift Observatory} \citep[\emph{Swift;}][]{Gehrels04} and the Gamma-Ray Burst Monitor (GBM, \citealt{Meegan2009}) onboard the \fermi spacecraft, revealed multiple clusters within the GRB population, suggesting the tantalizing possibility that some clusters correspond to specific progenitor types.

Ultimately, measurements that can definitively identify the progenitor type will be required to verify the robustness of any prompt emission-based classification schemes. For collapsars, this is the identification of a supernova \citep[SN, e.g.][]{Galama98,Hjorth03,Stanek03}, the smoking gun of a collapsing star. Recent examples of this include the identification of SNe accompanying the nominally short GRB~200826A \citep{Ahumada21,Zhang21,Rossi_2022}, the shortest known collapsar event; the unusually soft GRB 201015A \citep{Patel23,Belkin24}; and the Fast X-ray Transient (FXT) EP240414a \citep{vanDalen24}, whose broad-lined Type Ic SN links it to GRB progenitor stars. For mergers, positive confirmation can come in the form of accompanying GW radiation \citep{Abbott17_GRB} or kilonovae \citep{Berger13,Tanvir13,Yang15,Jin15,Jin16,Kasliwal17,Villar17,Gompertz18,Jin18,Troja18,Eyles19,Lamb19,Troja19,Jin20,Fong21,O'Connor21,Mei22,Troja2022,Yang2022,Levan24,Yang_2024}, but also from the definitive exclusion of SNe. A recent notable example of this was GRB 191019A, which may be the first example of a merger formed through dynamical capture \citep{Levan23, Stratta2025}, possibly in the accretion disc of an active galactic nucleus \citep{Lazzati23}.

Historically, the exclusion of a SN associated with the nearby ($z = 0.125$) long GRB 060614 to limits $100\times$ fainter than any previously known example \citep{Fynbo06,Gal-Yam06,Gehrels06} led to the recognition of the sub-population of so-called extended emission (EE) GRBs, of which 211211A and 230307A are members \citep{Gompertz23}. EE events are characterized by short, hard `spikes' of gamma-ray emission with typical durations of less than two seconds, followed by lower luminosity and spectrally softer gamma-ray emission that can last for tens to hundreds of seconds \citep{Norris06,Norris10}. The initial spikes are well matched to classic short GRBs, while the formal duration of EE GRBs extends well into the long GRB range. The ratio of the energy contained in the spike and the EE has been shown to be highly variable \citep{Perley09}, suggesting a continuum running from short GRBs with no EE through to rare examples of EE-dominated events like GRBs 080503 \citep{Perley09} and 191019A \citep{Levan23, Stratta2025}. EE is often attributed to the spin-down of a newly formed magnetar \citep{Metzger11,Bucciantini12,Gompertz13,Gompertz14} or fallback accretion processes \citep[e.g.][]{Rosswog07}, potentially due to a neutron star - black hole binary system \citep{Troja08,Desai19,Gompertz20}.

\thisgrb, detected at a redshift of $z = 2.681$ \citep{GCN_redshift}, emerges as a pivotal case in this evolving paradigm. 
Its light curve morphology, with an initial hard spike lasting $\sim 1.5$ s followed by weaker emission extending to $\sim 64$ s \citep{GCN_fermi}, is consistent with a short GRB with potential EE, positioning it as a candidate for a compact binary merger \citep{GCN_fermi, GCN_BAT-GUANO}. If confirmed as a merger-driven event, \thisgrb would be the most distant such GRB observed to date, offering a unique probe of neutron star merger rates in the early Universe. This is particularly significant for understanding \textit{r}-process nucleosynthesis, as standard merger models predict long delay times (hundreds of millions of years) that challenge their ability to enrich the early Universe with heavy elements \citep{Hotokezaka_2018, Cote_2019, Skinner_2024}. Confirming GRB 241105A as a higher redshift merger could support models for rapid binary evolution or strong natal kicks, enabling faster merger channels \citep{Belczynski06, Tauris13, Beniamini19, Belczynski02, OShaughnessy08}. As such, it would provide crucial observational evidence for the rapid production of heavy elements, representing a significant step toward solving the long-standing puzzle of chemical enrichment in the early Universe. 
Conversely, identifying it as a collapsar would cast further doubt on EE-like phenomenology as a reliable discriminant of progenitor type, deepening the GRB classification crisis. In both scenarios, this burst serves as a critical test case for GRB classification and rapid channels for r-process nucleosynthesis, making its study essential for advancing our understanding of GRB progenitors.

In this paper, we perform a multi-wavelength study of \thisgrb to probe its origin. The structure of this paper is as follows. In Section~\ref{sec:observations}, we present multiwavelength observations and data analysis of \thisgrb. Section~\ref{sec:prompt_diagnostics} examines the prompt emission characteristics of the burst. Section~\ref{sec:model} presents the afterglow modelling using multiband data. Section~\ref{sec:host} discusses the properties of the host galaxy, including SED fitting and afterglow offset analysis based on the James Webb Space Telescope (JWST) imaging. Finally, in Section~\ref{sec:discussion}, we discuss the progenitor scenario for the burst based on its prompt, afterglow and host properties. We assume a flat universe ($\rm \Omega_{k}$ = 0) with the cosmological parameters $\rm H_{0}$ = 70 km $\rm sec^{-1}$ $\rm Mpc^{-1}$, and the density parameters $\rm \Omega_{\Lambda}= 0.7$, and $\rm \Omega_m= 0.3$.

\section{Observations and data Reduction}
\label{sec:observations}
\thisgrb triggered the \gbm flight software at 16:06:04.66 UTC ($T_{0}$), 752515569 MET (trigger id: bn 241105671), and was identified as a short burst featuring a short spike followed by a weak EE. \gbm distributed an automated localization through the General Coordinates Network. The burst was also detected by \bat  \citep{GCN_BAT-GUANO}, Konus-Wind \citep{GCN_Konus_Wind}, and \grm \citep{svom_gcn}, with all missions confirming the presence of the weak emission following the main burst.
The burst triggered an extensive multi-wavelength campaign spanning gamma-ray, ultraviolet (UV), optical, near-infrared (NIR), and radio wavelengths.
This section outlines the observational data collected across these bands and the methodologies employed to analyze the event.

\subsection{High Energy Observations}
\label{subsec:gamma}

\subsubsection{{\it Fermi}/{\rm GBM}}
\label{subsec:fermi/gbm}
We extracted the Time-tagged Event (\texttt{TTE}) data for the event, and performed analysis using the {\tt GBM Data Tools} \footnote{\url{https://fermi.gsfc.nasa.gov/ssc/data/analysis/gbm/}}  Python library \citep{GDT-Fermi}.
The left panel of Figure \ref{fig:lc_241105A} shows the combined \gbm light curve for the Sodium-Iodide (NaI) detectors Na and Nb, along with the B1 Bismuth-Germanate (BGO) detector, which detects higher energy photons up to 40 MeV. The 2 NaI detectors were combined based on their source angle relative to the burst. Due to the relationship between detector effective area and source angle \citep{bissaldi2009ground}, detectors with source angles $\gtrsim 60^{\circ}$ are typically not used in temporal and spectral analyses. The burst also triggered detector N6. However, this detector would have been occulted by the spacecraft, and the initial pulse is not visible in the TTEs. Therefore, it is not included here. The background in the surrounding intervals was fit with a first order polynomial and is shown in red in the left panel of Figure \ref{fig:lc_241105A}. 
We estimated the \tninty for the burst using the GBM TTE and {\tt RMfit}\footnote{\url{https://fermi.gsfc.nasa.gov/ssc/data/analysis/rmfit}} software, in the 50–300 keV band. The \tninty duration is $69.0\pm13.5$~s, placing it well into the regime of long-duration GRBs. However, this duration includes a short, initial hard spike lasting $\sim$1.5\,s, followed by a prolonged weaker emission of around $\sim$64\,s. We divided the event in two emission episodes;  episode 1 from  $T_{0}$-0.256~s to $T_{0}$+1.28~s and episode 2 from $T_{0}$+1.28~s to $T_{0}$+\gbmTninty~s. We extracted the spectra for these two episodes using {\tt GBM Data Tools}. For this, we used the TTE data in the energy channels in the range of 8-900~keV for NaI detectors (Na, Nb) and 0.3-35~MeV for BGO detectors (B1). 

\subsubsection{\it{Swift}/{\rm BAT}}
\bat did not trigger on \thisgrb in real time. However, the \emph{Fermi} notice triggered the Gamma-ray Urgent Archiver for Novel Opportunities \citep[GUANO;][]{Tohuvavohu20}, which prompted the BAT event data between $T_{0}$-50s and $T_{0}$+150s to be saved and delivered to the ground. Ground analysis of these data revealed a detection of the GRB \citep[see][]{GCN_BAT-GUANO}.

BAT data were downloaded from the UK \emph{Swift} Science Data Centre \citep[UKSSDC;][]{Evans07,Evans09}. Per the header information in the event file, a \emph{Swift} clock correction of $-34.4815$s was applied to the photon arrival times to make them consistent with the data from the other high-energy satellites. The event data were processed using the standard High Energy Astrophysics Software \citep[HEASoft;][]{HEASoft} tools to apply mask weighting and a ray tracing solution. Spectra were created with the {\sc batbinevt} routine, and response matrices with the {\sc batdrmgen} script. The \emph{Swift}/BAT light curve of GRB 241105A is shown in the middle panel of Figure~\ref{fig:lc_241105A}.

\subsubsection{{\it SVOM}/{\rm GRM}}
The Space-based multiband astronomical Variable Objects Monitor \citep[\svom;][]{Cordier15,Wei2016,SVOM_ref} carries two wide-field high-energy instruments: a coded-mask gamma-ray imager \citep[ECLAIRs;][]{Godet14} and a Gamma-Ray Monitor \citep[GRM;][]{Dong10,Wen21}; and two narrow field telescopes: a Microchannel X-ray Telescope \citep[MXT;][]{Gotz23}, and a Visible-band Telescope \citep[VT;][]{Fan20}. \grm was triggered in-flight by \thisgrb (\svom burst-id: sb24110502) at 2024-11-05T16:06:05 UT by GRD01 and GRD03 \citep{svom_gcn}. The energy of GRM event data were calibrated using the GRM CALDB \footnote{\url{https://grm.ihep.ac.cn/CALDB.jhtml}}. The right panel of Figure~\ref{fig:lc_241105A} shows the \grm light curve combined with all three GRDs in different energy bands. It shows that the burst consists of a short single pulse with a duration of about 2 s and a weaker EE with a duration of about 50~s. The spectra were created within the same time intervals as \gbm. The background of \grm is estimated by fitting the data from $T_0-50$~s to $T_0-10$~s and $T_0+150$~s to $T_0+200$~s with first order polynomials. The response matrices were created from GRM CALDB. A joint gamma-ray spectral analysis incorporating \grm, \gbm, and \bat is performed later in Section~\ref{joint_spectral_fitting}.

\begin{table}
\centering
\caption{Photometric observations of the afterglow of \thisgrb. Magnitudes are in the AB system and are not corrected for Galactic extinction.}
\label{tab:photometric_data}
\begin{threeparttable}
\begin{tabular}{ccccl}
\toprule
$\Delta t$ (hours) & Filter & Magnitude (AB) & Telescope \\
\midrule
0.48  & $L$     & $17.21 \pm 0.01$           & GOTO \\
9.90  & $r$     & $19.20 \pm 0.20$           & BOOTES-7$^{\rm a}$ \\
11.08 & $R$     & $19.51 \pm 0.02$           & VLT/FORS2 \\
14.36 & $R$     & $19.88 \pm 0.03$           & SVOM/VT \\
16.04 & $R$     & $20.22 \pm 0.03$           & SVOM/VT \\
17.72 & $R$     & $20.68 \pm 0.04$           & SVOM/VT \\
19.37 & $R$     & $20.84 \pm 0.05$           & SVOM/VT \\
35.50 & $r$     & $22.66 \pm 0.10$           & ePESSTO+NTT \\
51.76 & $R$     & $22.33 \pm 0.15$           & SVOM/VT \\
14.36 & $B$     & $20.79 \pm 0.05$           & SVOM/VT \\
16.04 & $B$     & $21.21 \pm 0.06$           & SVOM/VT \\
17.72 & $B$     & $21.57 \pm 0.07$           & SVOM/VT \\
19.37 & $B$     & $21.52 \pm 0.06$           & SVOM/VT \\
51.76 & $B$     & $>23.7$                    & SVOM/VT \\
14.68 & white   & $22.24^{+0.11}_{-0.10}$    & \emph{Swift}/UVOT \\
16.84 & $o$     & $19.65 \pm 0.14$           & ATLAS \\
27.40 & white   & $23.73^{+0.53}_{-0.36}$    & \emph{Swift}/UVOT \\
35.61 & $z$     & $22.88 \pm 0.25$           & ePESSTO+NTT \\
68.44 & $L$     & $>19.81$                   & GOTO \\
\bottomrule
\end{tabular}
\begin{tablenotes}
\item \textbf{Note.} $^{\rm a}$\citet{GCN_BOOTES}.
\end{tablenotes}
\end{threeparttable}
\end{table}

\begin{figure*}
    \centering
    \includegraphics[width=0.47\linewidth]{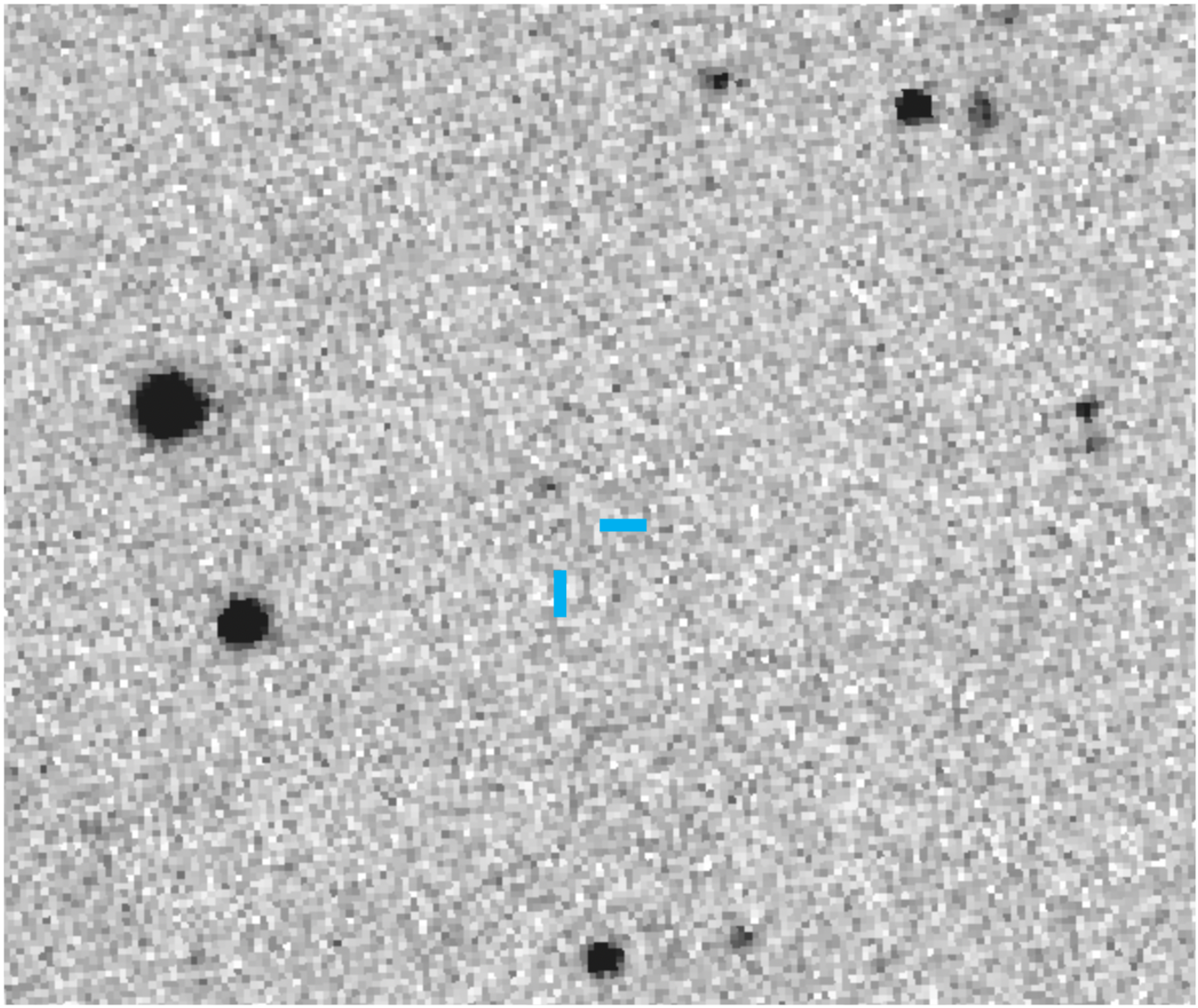}
    \includegraphics[width=0.47\linewidth]{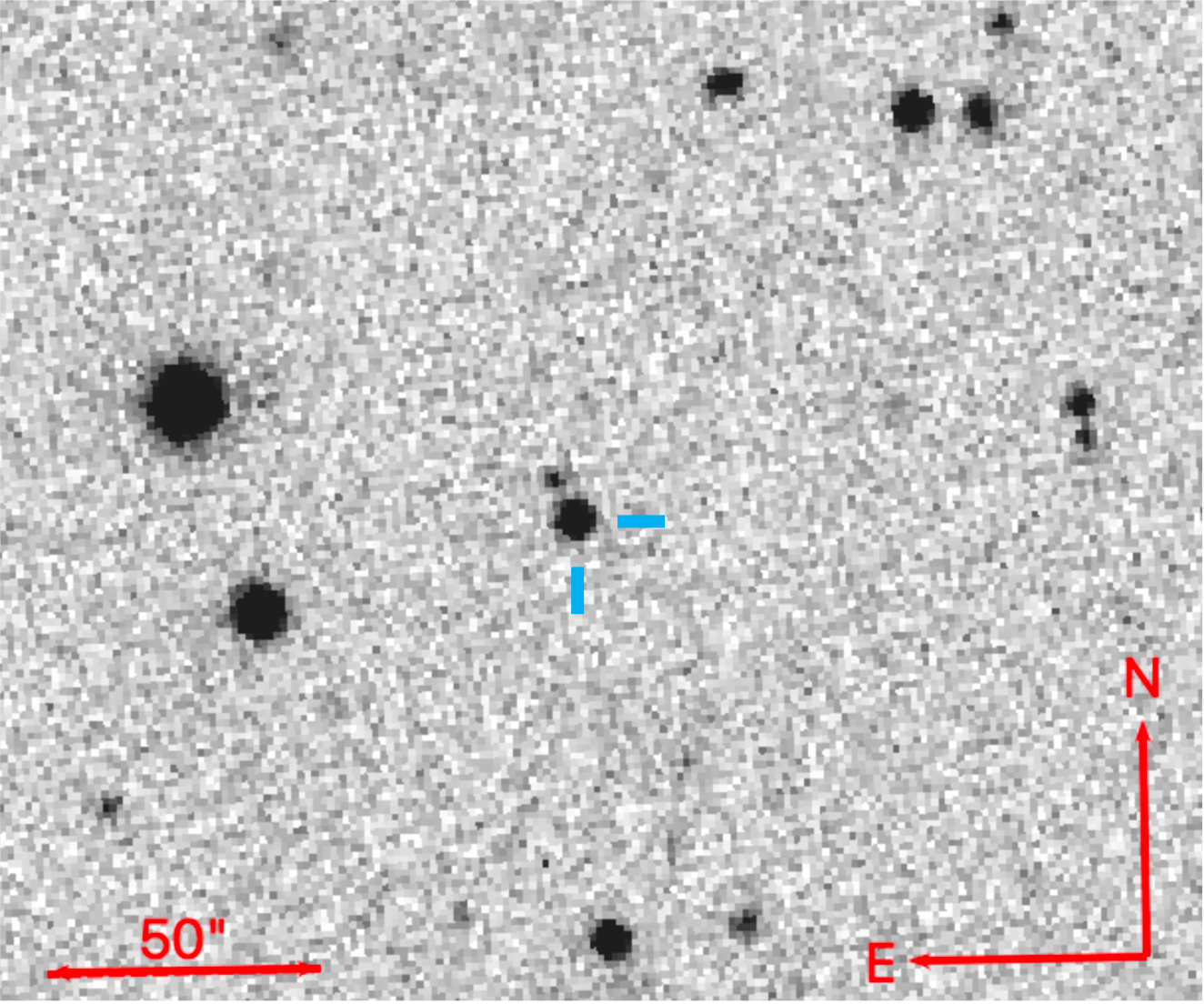}
    \caption{\textbf{\emph Left:}  A pre-trigger GOTO image of the field taken as part of the all-sky survey at 10:48:54 UT ($T_{0}$-5.29~hr), showing no source at the afterglow position.  
    \textbf{\emph Right:} The GOTO discovery image of the afterglow of \thisgrb, taken at 16:34:35 UT on 2024-11-05 ($T_{0}$+0.48~hr), shows the transient marked at the burst location. }
    \label{fig:GOTO_AG}
\end{figure*}
\subsection{Optical/UV Observations}
Photometric and spectroscopic follow-up of GRB 241105A was conducted across multiple facilities and filters, with observations spanning the ultraviolet to infrared wavelengths. All measured magnitudes and upper limits are compiled in Table~\ref{tab:photometric_data}.

\subsubsection{{\rm GOTO} afterglow discovery}
The Gravitational-wave Optical Transient Observer \citep[GOTO;][]{Dyer2020,Steeghs2022,Dyer2024} responded autonomously \citep{Dyer20b} to the \emph{Fermi} GCN notice \citep{GCN_fermi} from its southern site at Siding Spring Observatory in New South Wales, Australia. Across 10 unique pointings, GOTO-South tiled 277.9 square degrees within the \emph{Fermi}/GBM 90\% localisation region between 2024-11-05 16:19:10 UT (+0.22 hours post-trigger) to 2024-11-05 17:42:50 UT (+1.61 hours post-trigger), achieving a total coverage of $84.3$\% of the 2D localisation probability. Each pointing consisted of 4 $\times$ 90~s exposures in GOTO $L-$band (400–700 nm). Over the course of this imaging campaign, 104 images were captured with an average 5-sigma depth of $L = 20.4$ mag.

The data were processed promptly upon acquisition using the GOTO pipeline (Lyman et al. in prep.), incorporating difference imaging against deeper template observations of the same pointings to identify transient sources. An automated ML classifier \citep{Killestein2021} was employed to initially filter source candidates, which were subsequently cross-matched with contextual and minor planet catalogs to eliminate false positives. Any candidates passing these automated checks underwent real-time human vetting, ensuring rigorous verification of potential transient events.

After filtering out galactic variables and other false positives, a new optical source (internal name GOTO24ibf; IAU designation AT~2024aaon) was identified in images taken at 2024-11-05 16:34:35 (+0.48 hours post-trigger) at J2000 coordinates $\alpha = 04^{\mathrm{h}}24^{\mathrm{m}}59.00^{\mathrm{s}}, \ \delta = -49^{\circ}45^{\prime}09.33^{\prime\prime}$ \citep{GCN_goto}. 
The source was not present in earlier GOTO observations taken on 2024-11-05 at 10:48:54 UT, 5.29 hours before the GRB trigger, with a 3-sigma limiting magnitude of $L > 20.6$ (see Figure~\ref{fig:GOTO_AG}). Furthermore, no evidence of the source was found in archival data from the Asteroid Terrestrial-impact Last Alert System (ATLAS; \citealt{Tonry2018}) forced photometry server \citep{Shingles2021}, supporting its classification as a new transient source. The GOTO discovery refined the localization precision of \thisgrb from hundreds of square degrees to sub-arcsecond, enabling subsequent multi-wavelength observations that ultimately confirmed GOTO24ibf / AT~2024aaon as the afterglow \citep{GCN_redshift,GCN_VLT, GCN_Gemini/South, GCN_SVOM/VT, GCN_NTT, GCN_UVOT, GCN_BOOTES, GCN_XRT, GCN_ATCA}. Magnitudes and upper limits in the GOTO $L-$band were derived using forced photometry performed through the GOTO Lightcurve Service (Jarvis et al., in prep.).

\subsubsection{{\rm VLT/FORS2} spectroscopy and redshift determination}

We observed the optical counterpart \citep{GCN_goto} of GRB 241105A using the FOcal Reducer and low-dispersion Spectrograph (FORS2) mounted on the Very Large Telescope (VLT) Unit Telescope 1 (UT1; Antu, \citealt{Appenzeller1998}). The instrument was equipped with grisms 300V (without an order-sorting filter) and 300I (with the OG590 filter). We started observations on 2024-11-06 at 03:11:31 UT with an exposure time of 600\,s per grism. The optical counterpart was clearly detected in the acquisition image \citep{Izzo2024}. The magnitude is calibrated using the SkyMapper catalog and is tabulated in Table~\ref{tab:photometric_data}.

\begin{figure*}
    \centering
    \includegraphics[scale=0.7]{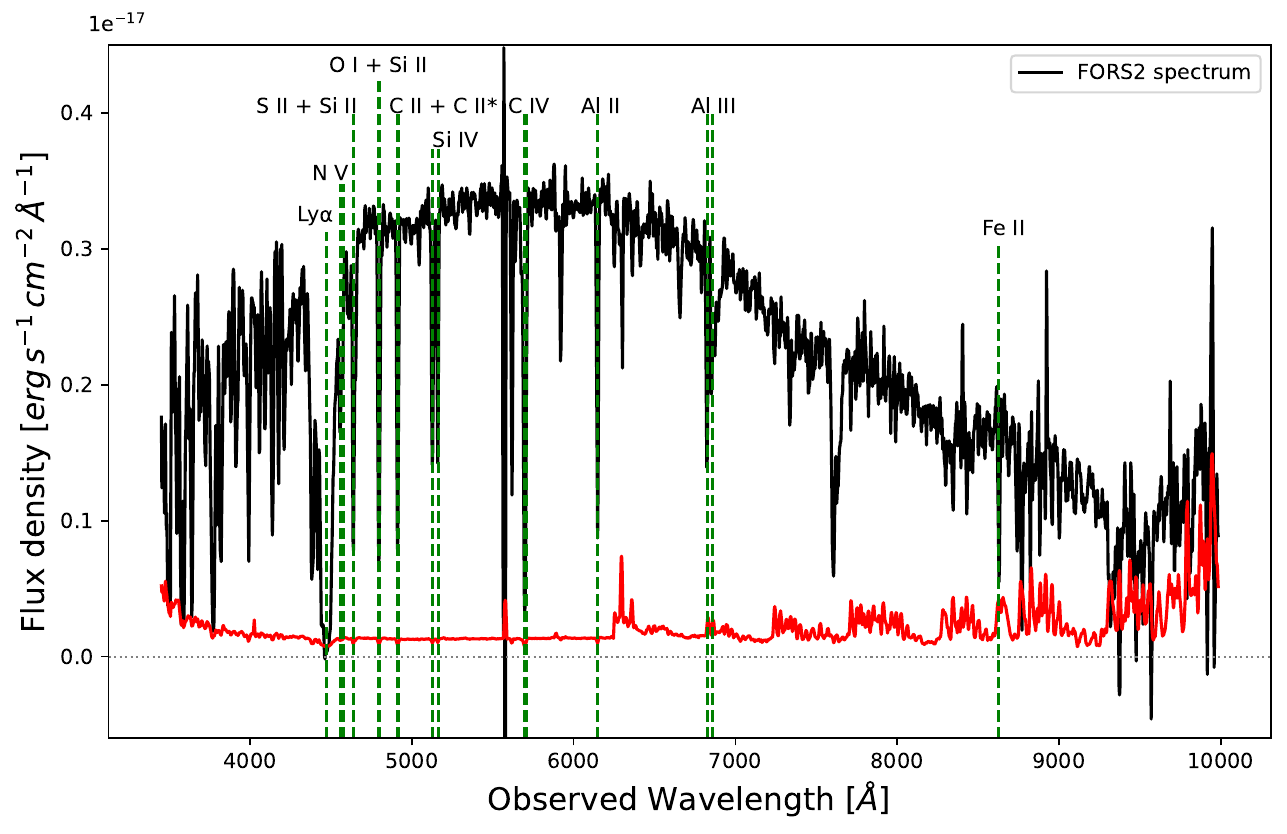}
    \caption{VLT/FORS2 optical afterglow spectrum of GRB\,241105A at redshift $z=2.681$. Data are in black, the error spectrum is in red, and the horizontal dotted line in grey corresponds to $F_{\lambda}=0$. The green vertical lines correspond to the labeled absorption lines.}
    \label{full_spec}
\end{figure*}

The spectral data were reduced using standard procedures. 
The final combined spectrum is shown in Figure \ref{full_spec}. A continuum is detected over the entire wavelength range from 3500 to 10000 \AA\,and a trough due to Ly$\alpha$ absorption is visible at the blue end. We identify several metal absorption features such as \sii{}, \siii{}, \siii{*}{}, \oi{}, \cii{}, \cii{*}{}, \feii{}, \alii{}, \nv{}, \civ{}, \siiv{}, and \aliii{}. The line profiles of low-ionisation transitions show a single strong component centred at $z=2.681$ (see Figure~\ref{full_spec}). High-ionization absorption lines also show only one component but blue-shifted by $\sim-150$\,km~s$^{-1}$. 
Further details on the absorption lines analysis are discussed in Section~\ref{sec:host} and Appendix~\ref{sec:CoG}. With a secure spectroscopic redshift of $z=2.681$, \thisgrb may represent the most distant short GRB known to date; if its origin is confirmed to be a neutron star merger, it would be the farthest such event yet discovered, significantly extending the redshift frontier for compact-binary merger-driven GRBs.

\subsubsection{{\rm NTT}}
The advanced Public ESO Spectroscopic Survey for Transient Objects (ePESSTO+; \citealt{Smartt_2015}) started observing the field of GRB241105A with the ESO New Technology Telescope (NTT) at La Silla, equipped with the EFOSC2 instrument in imaging mode. Observations started on 2024-11-07 at 03:21:27 UT, using the $r$ and $z$ Gunn filters. 
The $r$ and $z$ images consisted of 5 exposures of 180~s each. The $z$ band images were dithered to enable the fringing correction prior to alignment and stacking using a 2$\times$2 re-binned ESO NTT-EFOSC2 fringing mask provided by ESO\footnote{https://www.eso.org/sci/facilities/lasilla/instruments/efosc/inst/fringing.html}. 
Magnitudes were calibrated against the ATLAS-REFCAT2 catalogue \citep{ATLASrefcat}. The afterglow was clearly detected in $r$ and $z$ bands \citep{GCN_NTT}, and the magnitudes are reported in Table~\ref{tab:photometric_data}.

\subsubsection{\svom/{\rm VT}}
\svom/\emph{VT} has an effective aperture of 43 cm and a field of view of 26$\times$26 square arcminutes, giving a pixel scale of $0.76^{\prime\prime}$.
It conducts observations with two channels, VT\_B and VT\_R, simultaneously, covering wavelengths of 400-650~nm and 650-1000~nm, respectively.
Detailed information on VT will be presented in Qiu et al. (2025, in preparation). 
During the commissioning phase, VT observed \thisgrb using the Target Of Opportunity (ToO) mode on 2024-11-06 and 2024-11-07.
Five orbits covering 7~hours were scheduled with the earliest observation starting on 2024-11-06, about 14.2~hours post trigger time.
The counterpart was clearly detected in all stacked images in each orbit \citep{GCN_SVOM/VT}.
On 2024-11-07, the counterpart was marginally detected in R-band stacked images at about 51.7~hours post trigger time and undetected in B-band stacked images with a total exposure time of 6800~s.
For all observations, individual exposures were set to 20~s per frame.
All data were processed in a standard manner, including zero correction, dark correction, and flat-field correction. 
After pre-processing, the images for each band obtained during each observation were stacked to increase the signal to noise ratio.
Photometric measurements, calibrated in the AB magnitude system, are presented in Table~\ref{tab:photometric_data}.

\subsubsection{Swift/{\rm UVOT}}
The \swift Ultra-Violet and Optical Telescope (UVOT, \citealt{Roming2005}) began observations of the field of \thisgrb around $10.6$ hours 
after the \gbm trigger \citep{GCN_UVOT}. Observations were taken with the $white$ filter only. A faint fading source was detected at the afterglow position \citep{GCN_UVOT}. Source counts were extracted from the UVOT image mode data using a source region of $5^{\prime\prime}$ radius. Background counts were extracted using four circular regions of radius $10^{\prime\prime}$ located in source-free regions near the GRB. The count rates were obtained from the co-added images using the {\it Swift} tool \texttt{uvotsource}. They were converted to magnitudes using the UVOT photometric zero points
\citep{bre11,poole}. 

ß\begin{table}
    \centering
    \caption{JWST/NIRCam AB magnitudes of the host galaxy of \thisgrb\ measured in two apertures.}
    \label{tab:jwst_photometry}
    \begin{threeparttable}
    \begin{tabular}{lcc}
        \toprule
        Filter & 0.1\arcsec\ Aperture (mag) & 1\arcsec\ Aperture (mag) \\
        \midrule
        F070W & $24.52 \pm 0.006$ & $23.28 \pm 0.15$ \\
        F115W & $24.30 \pm 0.004$ & $23.14 \pm 0.05$ \\
        F150W & $24.89 \pm 0.005$ & $22.87 \pm 0.05$ \\
        F277W & $23.87 \pm 0.001$ & $22.54 \pm 0.03$ \\
        F356W & $24.06 \pm 0.002$ & $22.67 \pm 0.04$ \\
        F444W & $23.73 \pm 0.002$ & $22.54 \pm 0.03$ \\
        \bottomrule
    \end{tabular}
    \end{threeparttable}
\end{table}

\begin{figure}
    \centering
    \includegraphics[width=\columnwidth]{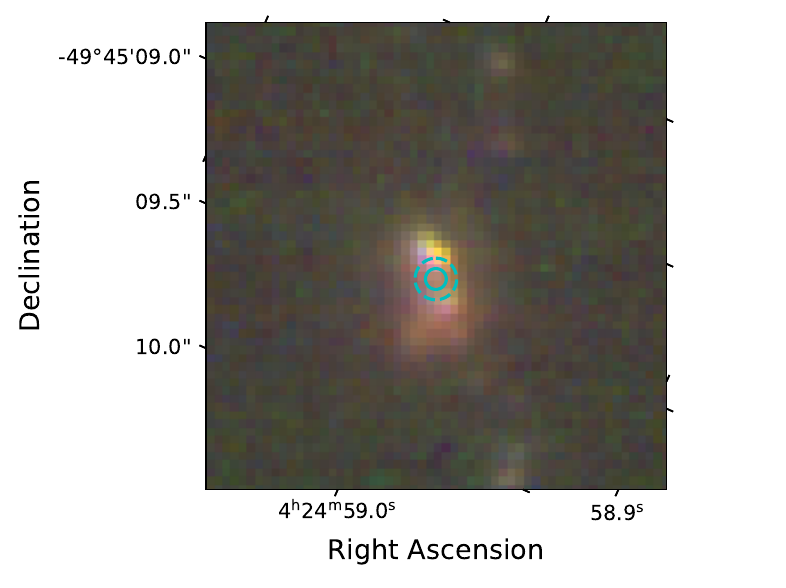}
    \caption{The position of GRB241105A overlaid on a composite JWST NIRCam image combining filters F070W, F115W, and F150W, as described in Section~\ref{sec:offset}. The disturbed/irregular morphology of the host is clearly visible. One (solid circle, 36\,mas) and two (dashed circle, 72\,mas) sigma uncertainties on the transient position are shown.}
    \label{fig:location}
\end{figure}
\subsection{JWST Observations}
\label{obs:jwst} 
We conducted target-of-opportunity observations with JWST under Director's Discretionary Time (ID: 9228, PI: Dimple). 
Observations were taken with the Near InfraRed Camera \citep[NIRCam;][]{Rieke22} in six wide filters (F070W, F115W, F150W, F277W, F356W and F444W), beginning on 2024-12-22 at 17:26:00 UT—approximately 47 days after the burst (\(\sim12.7\) days in the GRB rest frame), close to the predicted observer‐frame peak of any accompanying supernova \citep[e.g.][]{Galama98,Patat01}. We detected a galaxy in all the observed bands, which archival images \citep{GCN_goto} had previously identified. The galaxy was resolved into at least two distinct extended regions separated by $\sim 0.2^{\prime\prime}$, as shown in Figure~\ref{fig:location}. We performed photometric analysis in a $0.1^{\prime\prime}$ aperture at the afterglow location. We obtained a magnitude of F277W(AB)\(=23.87 \pm 0.001\), approximately 2.4 magnitudes brighter than the anticipated SN 1998bw-like peak, which makes host subtraction essential to reveal any transient. In addition, we performed photometric analysis for the complete galaxy, which was later used for the host analysis. All photometric measurements are reported in Table~\ref{tab:jwst_photometry}.

\subsection{Radio Observations}
\begin{table}
    \centering
    \caption{Radio flux density measurements 
    of the radio afterglow of \thisgrb. All errors are $1\sigma$. The $3\sigma$ thresholds are 3 times the RMS and correspond to the upper limit for all non-detections.}
    \label{tab:atca}
    \begin{threeparttable}
    \begin{tabular}{lccc}
        \toprule
        Time post-burst & Frequency $\nu$ & Flux density & $3\sigma$ threshold \\
        (days) & (GHz) & ($\mu$Jy/beam) & ($\mu$Jy/beam) \\
        \midrule
        4.8  & 5.5  & --           & 54 \\
        4.8  & 9.0  & --           & 39 \\
        15.0 & 5.5  & --           & 36 \\
        15.0 & 9.0  & $115 \pm 43$ & 33 \\
        17.1 & 18.0 & $343 \pm 46$ & 96 \\
        22.1 & 5.5  & $47 \pm 45$  & 114 \\
        22.1 & 9.0  & $220 \pm 61$ & 78 \\
        22.1 & 18.0 & $571 \pm 154$& 180 \\
        \bottomrule
    \end{tabular}
    \end{threeparttable}
\end{table}

\subsubsection{ATCA}
Observations with the Australia Telescope Compact Array (ATCA) 4\,cm dual receiver (central frequencies of 5.5 and 9\,GHz both with a 2\,GHz bandwidth) commenced 4.8\,days post-burst under program C3204 (PI: Anderson). 
Following the standard techniques, the data were processed using {\sc Miriad} \citep{sault95} with PKS 1934-638 and PKS 0437-454 as the primary and gain calibrator, respectively.
While no radio counterpart was detected during this first epoch \citep{GCN_ATCA_1}, a follow-up observation 15 days post-burst detected the afterglow at 9\,GHz \citep{GCN_ATCA}. We then performed higher frequency follow-up with the 15\,mm dual receiver (central frequencies of 17 and 19\,GHz) at 17\,days post-burst, which was combined in the visibility plane to obtain a detection at 18\,GHz. A further observation was obtained 22\,days post-burst showing the afterglow had brightened 
at both 9 and 18\,GHz. 
The resulting flux densities and $3\sigma$ upper limits (3 times the RMS) can be found in Table~\ref{tab:atca}.

\begin{table*}
\centering
\caption{Spectral fitting results for GRB 241105A in different time intervals. Uncertainties correspond to 1$\sigma$ confidence intervals.}
\label{tab:grb241105a_spectra}
\renewcommand{\arraystretch}{1.4}
\begin{threeparttable}
\begin{tabular}{lcccccccc}
\toprule
Emission Region & Time Range (s) & Model & Flux\tnote{a} & $E_{\rm p}$ (keV) & $\alpha$ & $\beta$ & $E_{\rm iso}$\tnote{b} & $\chi^2/\mathrm{dof}$ \\
\midrule
Episode 1 & $-0.256$ to $1.28$ & Band         & $12.4 \pm 0.4$                 & $312.0^{+168.2}_{-134.1}$ & $1.15^{+0.12}_{-0.17}$ & $2.08^{+0.29}_{-0.15}$ & $3.20^{+0.10}_{-0.10}$  & 1.13 \\
Episode 2 & $1.28$ to $64.961$  & Cutoff PL    & $1.12^{+0.05}_{-0.06}$        & $3127.8^{+3015.7}_{-1750.3}$ & $1.25^{+0.11}_{-0.12}$ & --                    & $11.97^{+0.53}_{-0.64}$ & 1.79 \\
Whole burst & $-0.256$ to $64.961$ & Cutoff PL  & $1.43^{+0.05}_{-0.06}$        & $2381.6^{+2014.0}_{-1291.2}$ & $1.24^{+0.09}_{-0.11}$ & --                    & $15.65^{+0.55}_{-0.66}$ & 1.78 \\
\bottomrule
\end{tabular}
\begin{tablenotes}
\footnotesize
\item[a] Flux is given in units of $10^{-7}$ erg s$^{-1}$ cm$^{-2}$.
\item[b] Isotropic equivalent energy $E_{\rm iso}$ is in units of $10^{52}$ erg.
\end{tablenotes}
\end{threeparttable}
\end{table*}

\section{Prompt Emission Analysis}
\label{sec:prompt_diagnostics}
\subsection{Joint Spectral Analysis}
\label{joint_spectral_fitting}
We performed a joint fit of the \gbm, \bat, and \grm data to maximize the signal-to-noise ratio, allowing more robust constraints on spectral parameters. Spectra were extracted in the same time intervals as \gbm as defined in section~\ref{subsec:fermi/gbm}. The fitting was performed in {\sc xspec} v12.14.1 \citep{Arnaud96}. We fit the power-law, Comptonized, and Band function models and compare their goodness-of-fit statistics using Akaike's Information Criterion \citep{Akaike74}. A more complex model is preferred when $\Delta_{\rm AIC} \geq 4$, otherwise the simpler model is chosen. Free normalisations were included in the model fits to account for the relative effective areas between detectors on the different satellites. We found that the GRM detectors also required free normalisations relative to one another, as determined by $\Delta_{\rm AIC}$. We found that the inclusion of \grm for the Episode 2 and complete burst fits only added noise and resulted in a less constrained fit, so these epochs were fit with \emph{Fermi}/GBM and \emph{Swift}/BAT data only. The best-fitting models are presented in Table~\ref{tab:grb241105a_spectra}. Using these values, we computed the isotropic-equivalent energies and source-frame peak energies for both emission episodes and the whole burst, which are reported in Table~\ref{tab:grb241105a_spectra}.

\begin{figure}
    \centering
   \includegraphics[width=\columnwidth]{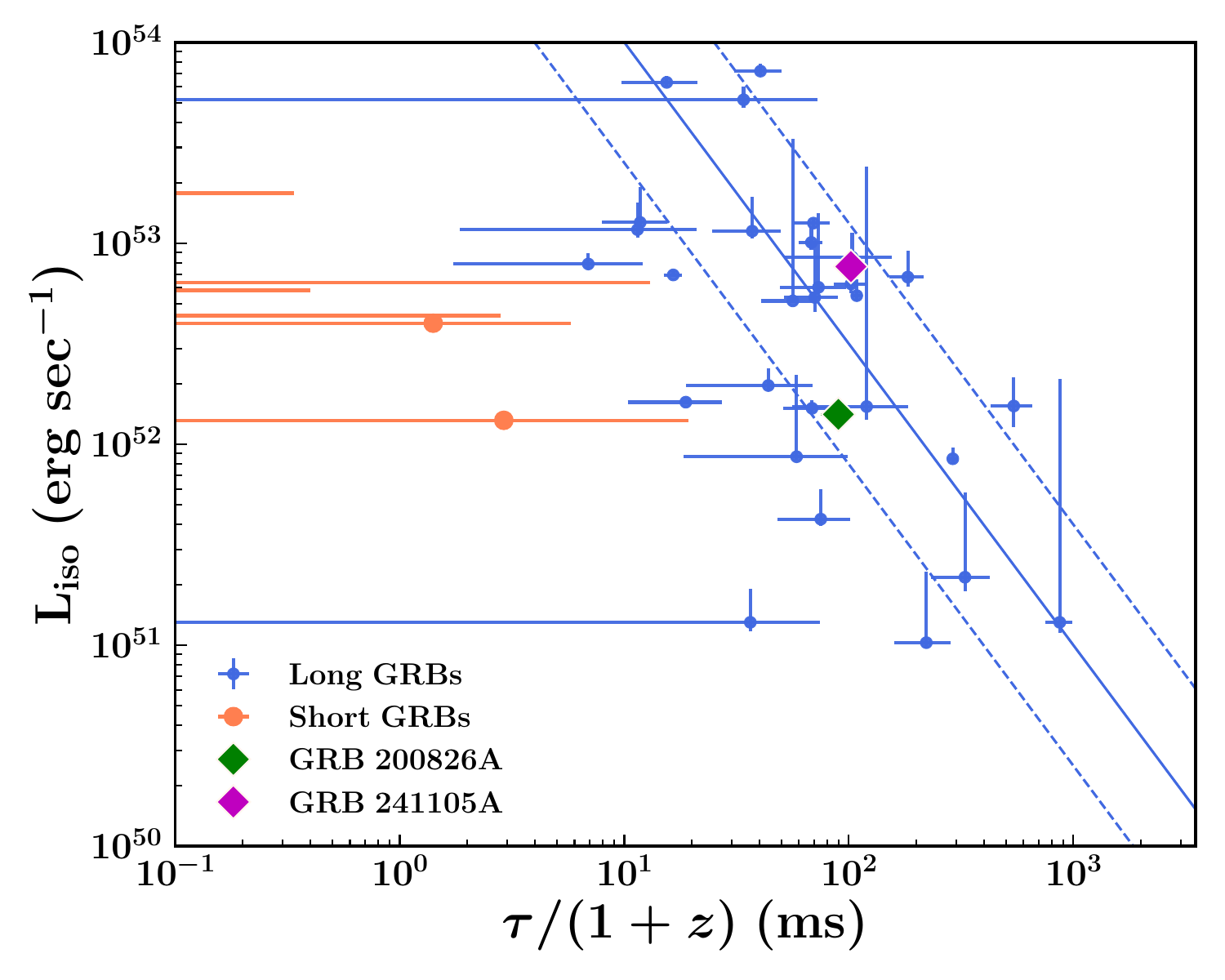}
    \caption{ Lag-luminosity correlation of GRBs using data from \citet{Ukwatta2010}. \thisgrb follows the lag-luminosity trend typical of long GRBs, unlike short GRBs.}
    \label{fig:lag}
\end{figure}
\begin{table*}
\centering
\begin{threeparttable}
\begin{tabular}{llccc}
\toprule
Instrument & Emission episode & Hardness Ratio & Minimum variability timescale (s) & Spectral l
ag (ms) \\
\midrule
\multirow{3}{*}{\textit{Swift}/BAT} 
& Episode 1 & $1.34 \pm 0.17$ & $0.50 \pm 0.17$ & $376 \pm 11$ \\
& Episode 2 & $1.60 \pm 0.25$ & $27.37 \pm 0.15$ & -- \\
& Full Burst & $1.58 \pm 0.29$ & -- & -- \\
\midrule
\multirow{3}{*}{\textit{Fermi}/GBM} 
& Episode 1 & $0.856 \pm 0.211$ & $0.31 \pm 0.09$ & $130.8 \pm 230.8$ \\
& Episode 2 & $0.354 \pm 0.328$ & $22.58 \pm 4.07$ & -- \\
& Full Burst & $0.625 \pm 0.034$ & -- & -- \\
\bottomrule
\end{tabular}
\caption{Prompt emission properties of \thisgrb for Episode 1, Episode 2, and the full burst, as measured from \bat and \fermi data.}
\label{tab:prompt_properties}
\end{threeparttable}
\end{table*}

\subsection{Spectral lag}
We computed the spectral lag, which refers to the delay in the arrival times of low-energy photons relative to high-energy photons. Long GRBs typically exhibit significant spectral lags, with delays of up to a few seconds in their light curves across different energy channels. In contrast, short GRBs generally show little or no spectral lag, \citep{Cheng_1995, Yi_2006}. For \bat data, we estimated the lag between the 15–25 keV and 50-100 keV light curves using the cross-correlation function (CCF) method \citep{Bernardini2015}. The lag is found to be $\tau_{\mathrm{CC}}= 376\pm11$ ms. 
Also, an anti-correlation between the bolometric peak luminosity and the spectral lag of GRBs has been identified by \citet{Norris02}, and subsequently confirmed by \citet{Gehrels06}, and \citet{Ukwatta2010}. Figure~\ref{fig:lag} shows the position of \thisgrb in the lag-luminosity plane. The burst lies within the 2$\sigma$ region of the correlation defined by long GRBs, as also seen previously for the short collapsar GRB~200826A \citep{Rossi_2022,Dimple2022}. For \gbm data, we computed the spectral lag between the 25–50 keV and 100–300 keV light curves. The lag is found to be $\tau_{\mathrm{CC}} = 130.8 \pm 230.8$\,ms, and is therefore too poorly constrained to be informative.

\begin{figure}
    \centering
    \includegraphics[width=\columnwidth]{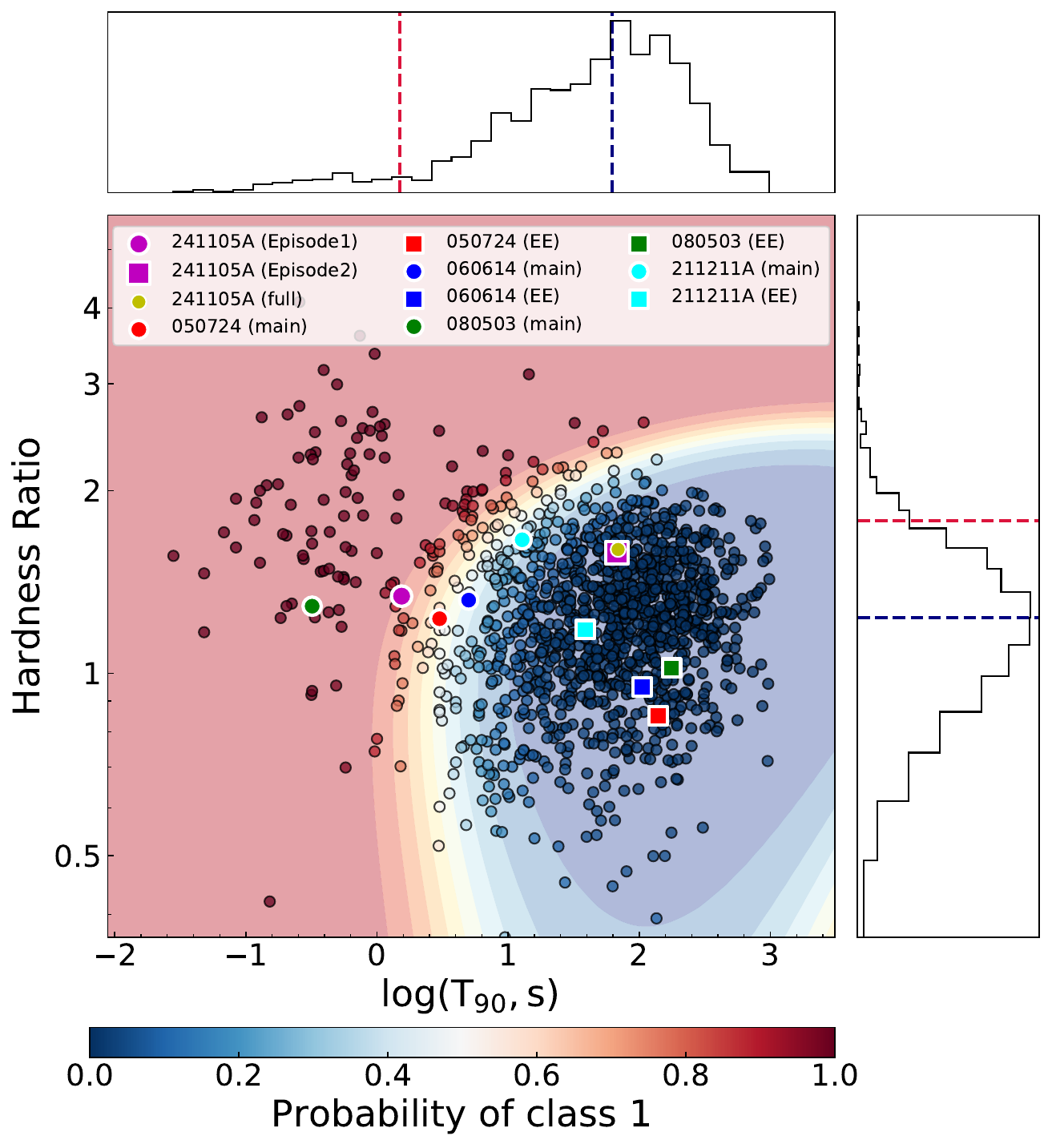}
    \caption{The positions of GRB 241105A (Episodes 1 and 2) are shown in the log (\tninty)-HR plane, overlaid on the \bat GRB population alongside three well-known short GRBs with extended emission and long mergers with extended emission. The background contours represent the probability density of two components identified by the BGMM, corresponding to the short-hard and long-soft GRB classes, with the colorbar indicating the probability of belonging to the short class. Marginal histograms show the distributions of \tninty and HR, with dashed lines marking the mean values for each class.}
    \label{fig:HR_T90}
\end{figure}
\subsection{Hardness Ratio}
We computed the hardness ratio (HR) as the fluence ratio between higher and lower energy bands, using data from both \bat and \gbm for both the emission episodes as mentioned in Section~\ref{subsec:gamma}.
Using \bat data, we calculated the HR between the 50–100 keV and 25-50 keV bands. For episode 1 (T$_0$--0.256 to T$_0$+1.28\,s), the HR is measured to be $1.34\pm0.17$. For episode 2, the HR is $1.60\pm0.25$, and for the whole burst duration, it is $1.58\pm0.20$.
We fitted a Bayesian Gaussian Mixture Model (BGMM) in the $T_{90}$ versus HR  distribution, using the \bat sample \citep{Lien2016}, to estimate the likelihood of the burst belonging to the short or long GRB classes. Figure~\ref{fig:HR_T90} shows the location of \thisgrb in the HR–\tninty plane relative to these EE GRBs. The background contours represent the probability density distribution of the two Gaussian components (dividing them into two classes, class 1 and class 2) derived from the fitted BGMM model. The probability of Episode 1 belonging to class 1 (which is closer to short GRBs) is $P(\mathrm{class \ 1}) = 0.886$, indicating a strong association with the short-hard GRB population. In contrast, Episode 2 yields $P(\mathrm{class \  1}) = 0.030$, aligning more closely with the long-soft GRB class. The whole emission has $P(\mathrm{class \  1}) = 0.032$, which also aligns with a long burst.

We also performed an HR analysis using GBM data, calculating the fluence ratio between the 50–300 keV and 10–50 keV bands. The HR for episode 1 is $0.856 \pm 0.211$, while for the whole burst it decreases to $0.625 \pm 0.034$. The HR for episode 2 alone is $0.354 \pm 0.328$, although this measurement is affected by larger uncertainties due to lower photon statistics.

Further, we examined whether this burst aligns with classical EE GRBs, characterized by a hard initial spike followed by softer, prolonged emission \citep{Norris06, Perley09}. We compared the HR of \thisgrb with three well-studied short GRBs exhibiting EE: GRB~050724 \citep{Barthelmy2005}, GRB~060614 \citep{Gehrels06}, GRB~080503 \citep{Perley09}, as shown in Figure~\ref{fig:HR_T90}. We also included the long merger GRB~211211A in our analysis, which also exhibits the characteristics of an EE burst. 
In typical EE events, the extended component is spectrally softer than the initial spike \citep[e.g.,][]{Kaneko2015}. However, in the case of \thisgrb, we find that the second episode is spectrally harder than the first, in contrast to the canonical soft EE. While the temporal structure superficially resembles EE GRBs, the distinct spectral behaviour suggests that Episode 2 may not represent a classical EE component.

\begin{figure}
    \centering
    \includegraphics[width=0.97\columnwidth]{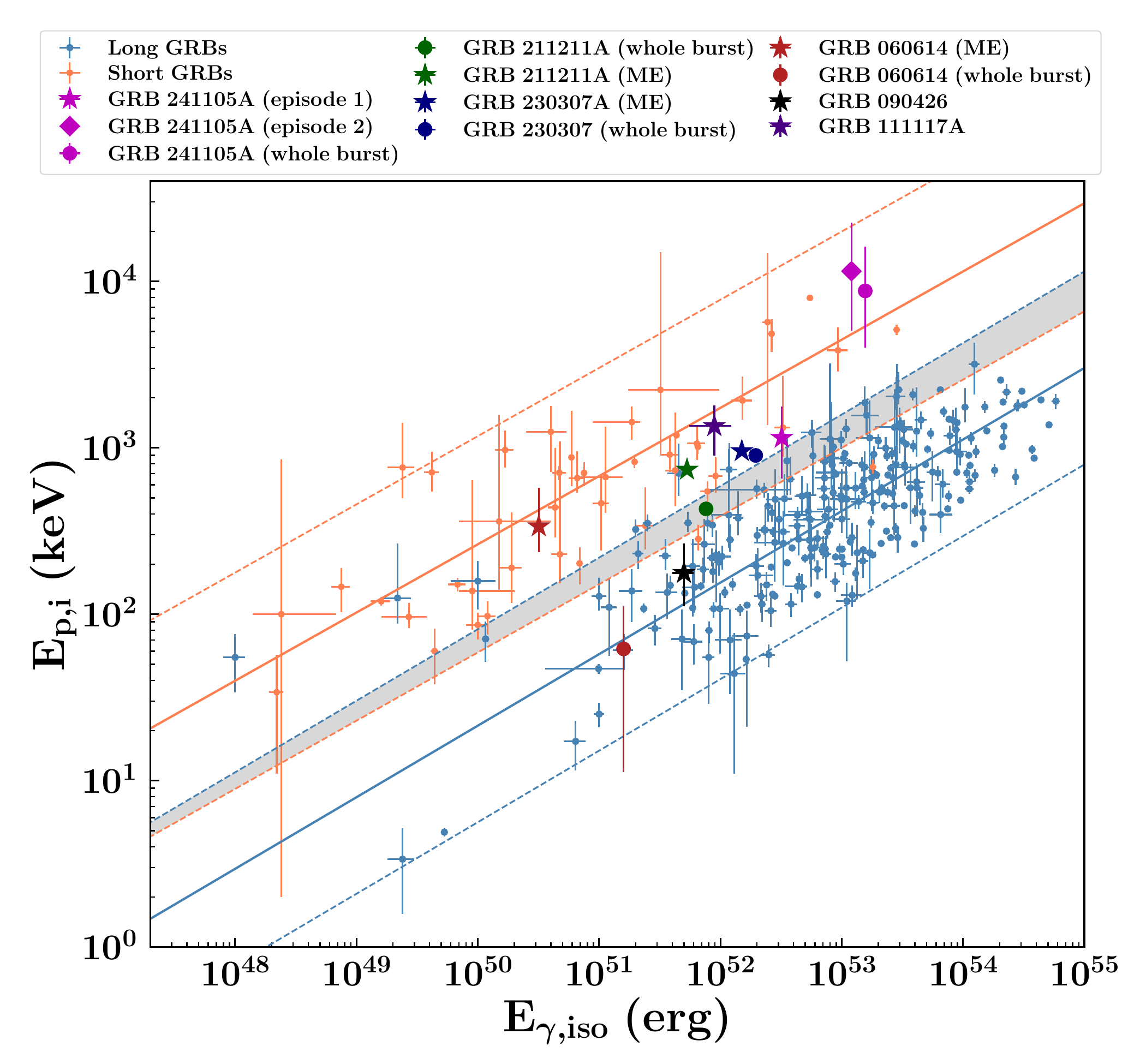}
    \caption{Location of \thisgrb on the Amati plane ($\rm E_{p,i}$ vs. $\rm E_{\gamma,\mathrm{iso}}$) along with short and long GRBs using the dataset from \citet{M2019}.}
    \label{fig:amati}
\end{figure}

\subsection{Location on the Amati plane}
Using the values derived in section~\ref{joint_spectral_fitting}, we placed episode 1, 2, and the whole burst in the Amati plane, alongside the short and long GRBs, using the dataset from \citet{M2019} as shown in Figure~\ref{fig:amati}. Episode 1 lies near the overlapping region of long and short GRBs. However, Episode 2 and the whole burst are located within the region typically occupied by short GRBs. It is important to note that this comparison involves a heterogeneous dataset compiled from multiple instruments and missions, each with differing sensitivities, calibrations, and analysis methodologies, which may introduce systematic uncertainties in the derived parameters. We also locate GRB 090426 and GRB 111117A, short GRBs at redshift $z>2$. Interestingly, GRB 090426 is located within the region typically occupied by long GRBs in the Amati plane. In addition, the main emissions (MEs) and whole emissions of GRB 060614, GRB 211211A, and GRB 230307A (compiled from \cite{Zhu_2022} and \cite{Peng_2024}) are also included in our sample, which are long mergers with EE, as confirmed either by the detection of a kilonova or the exclusion of a SN.

\subsection{ML Insights}
In addition to traditional classification methods, machine learning has become an indispensable tool in clustering GRBs, enabling the identification of underlying patterns and subpopulations within their complex datasets \citep{Jespersen20, Dimple23, Dimple24}. To investigate the properties of \thisgrb, we employed a machine learning pipeline that integrates Principal Component Analysis (PCA; \citealt{Hotelling1933}) with Uniform Manifold Approximation and Projection (UMAP; \citealt{McInnes2018}). This analysis utilised prompt emission light curves from the \gbm catalog across three energy bands—8–50 keV, 50–300 keV, and 300–1000 keV—with a temporal resolution of 16 ms. 
We preprocessed the light curves by normalising the fluence, aligning the start times, zero-padding to achieve uniform duration, and applying a Discrete-Time Fourier Transform to preserve temporal delay signatures as detailed in \citet{Dimple23, Dimple24}. 
We first used PCA to reduce the dimensionality of the data by transforming it into orthogonal principal components, retaining approximately 99\% of the total variance to capture salient features of the data while suppressing noise. Subsequently, UMAP is used to project the PCA-reduced data into a two-dimensional embedding space, preserving both local and global structures through a topological similarity matrix. We optimised UMAP’s performance using the hyperparameters \texttt{n\_neighbors}=25 and \texttt{min\_dist}=0.01, identified through iterative tuning. 

\begin{figure}
    \centering
    \includegraphics[width=\columnwidth]{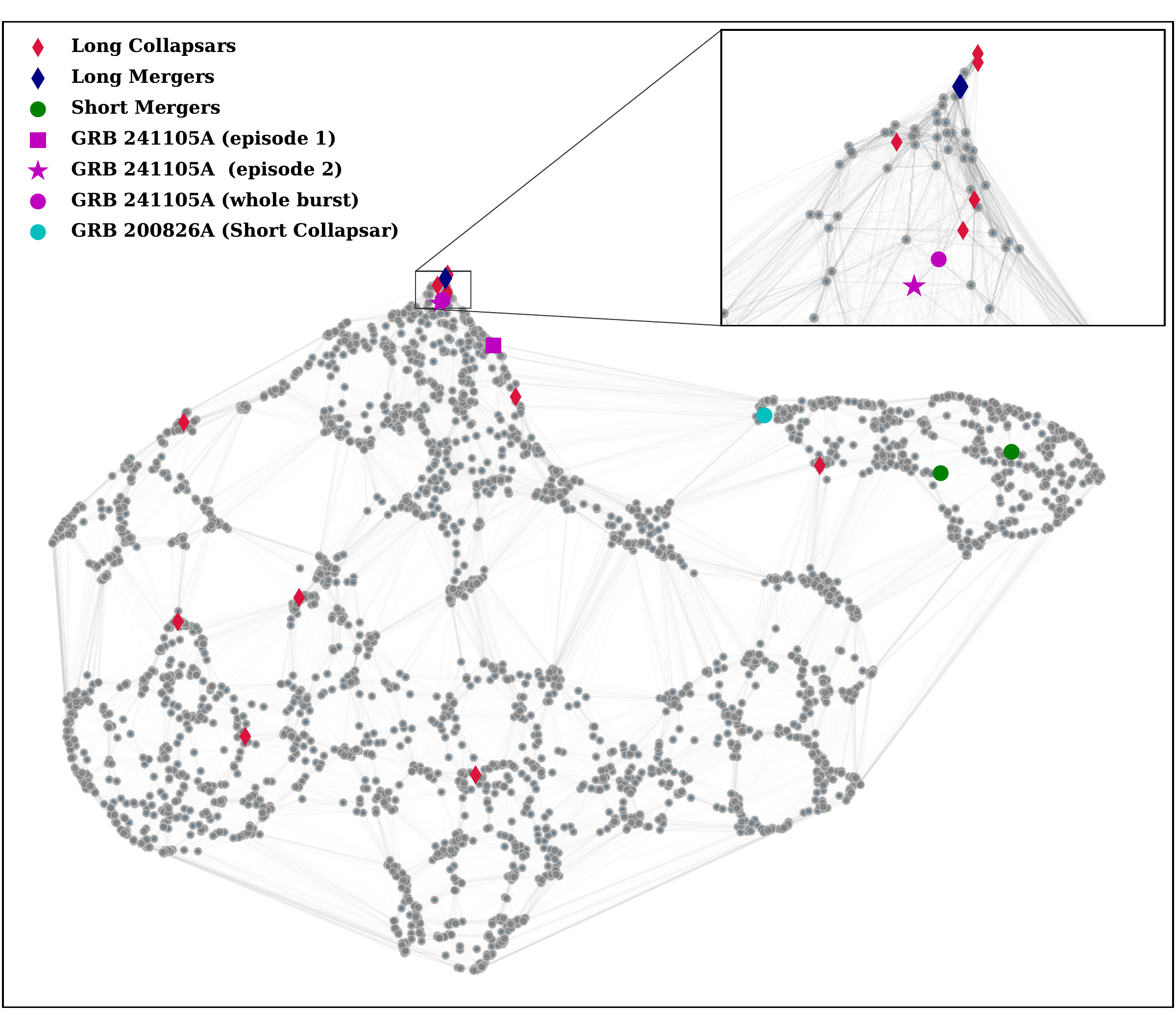}
    \caption{Two-dimensional embedding obtained using uniform manifold approximation and projection (UMAP) initialised by principal component analysis (PCA) obtained using \gbm light curves in different energy bands. The distances between points in the embedding give an idea of the similarity between the two events.}
    \label{fig:ML}
\end{figure}

Figure \ref{fig:ML} shows the connectivity map for \gbm bursts. In this map, GRBs lying close to each other suggest that they share similar properties in the high-dimensional feature space. As the input comprises prompt emission light curves, such similarity may not necessarily reflect progenitor type, but could instead indicate similarities in central engine behaviour or radiation mechanisms. The proximity of \thisgrb in the machine-learning embedding to both collapsars and long-duration mergers indicates similarities in their central engines (e.g., black hole accretion or magnetar spin-down) or emission mechanisms (e.g., prolonged jet activity or synchrotron radiation), rather than a common progenitor. This degeneracy necessitates multi-wavelength data to confirm the progenitor. In addition, we located both episodes separately on the map; it is interesting to see that even Episode 1 lies in the cluster surrounded by long GRBs. Both episodes and the whole burst lie far from short-duration mergers, indicating that if \thisgrb is merger-driven, it resembles long-duration mergers rather than short-duration mergers.

\subsection{Minimum Variability Timescale}
\begin{figure}
  \includegraphics[width=0.97\columnwidth]{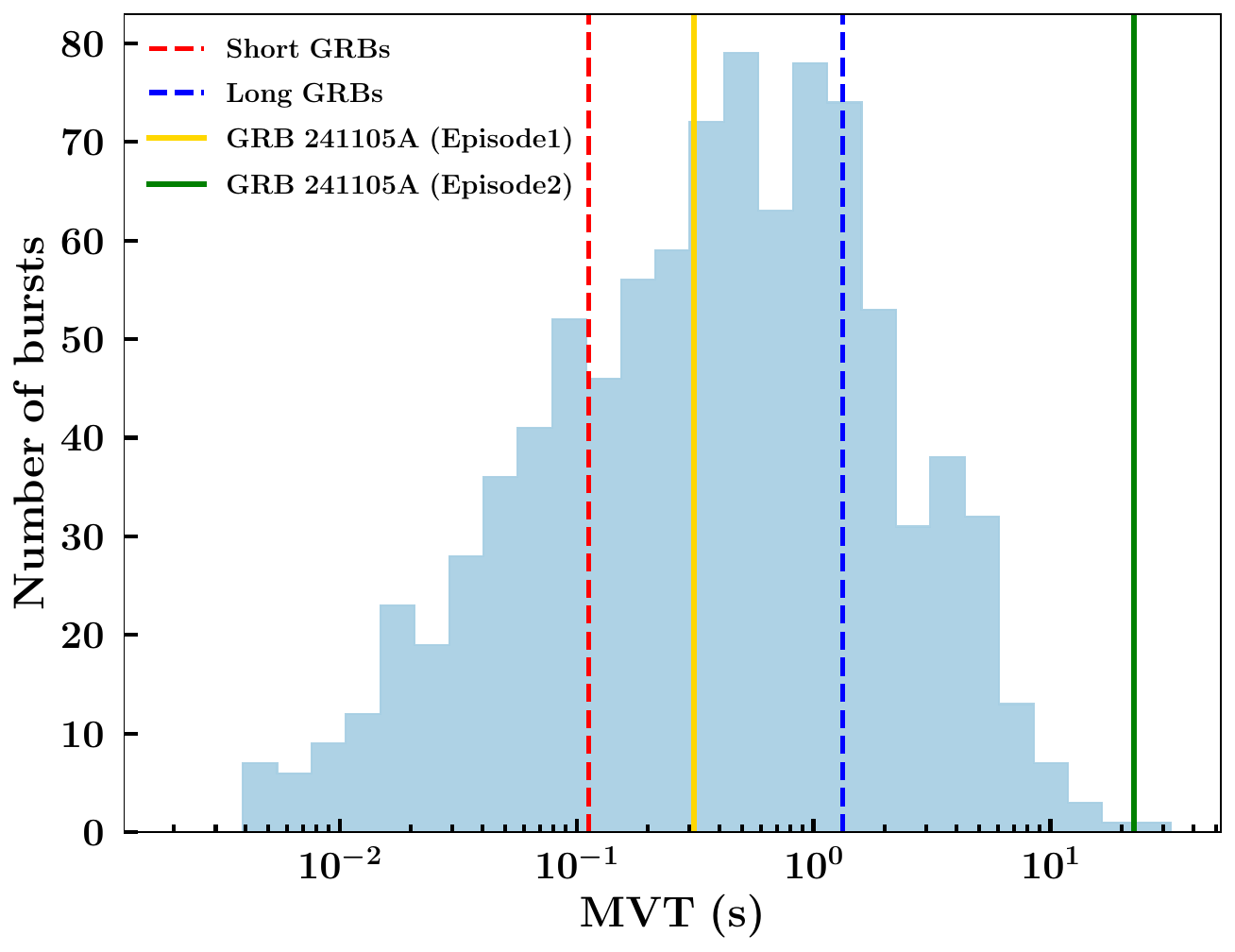}    
    \caption{MVT histogram using Fermi/GBM sample from \citet{Golkhou+15mvt}. The yellow and green lines represent the MVT for both the emission episodes. The dotted lines show the mean value of MVT for short (red) and long (blue) GRB populations. }
    \label{fig:MVT_hist}
\end{figure}

Minimum variability timescale (MVT), which traces the shortest resolvable timescales in the light curve, provides constraints on the size and dynamics of the emitting region and the central engine \citep{MacLachlan13, Golkhou+15mvt, Camisasca2023, Maccary2025}. 
We estimated the MVT following the method of \citet{Golkhou+15mvt}, using both \bat and \gbm data following \citet{Golkhou+15mvt}. For the \gbm data in the 8–900 keV energy range, Episode 1 exhibits a MVT $= 0.31 \pm 0.09$ s, while Episode 2 shows a significantly longer MVT of $22.58 \pm 4.07$ s. These values are illustrated in Figure~\ref{fig:MVT_hist}, which shows the MVT distribution for the \gbm sample from \citet{Golkhou+15mvt}. The yellow and green lines mark the MVTs for both episodes, respectively, while the red and blue dashed lines indicate the mean MVT values for the short and long GRB populations. The MVT of the first episode falls between the characteristic mean values for short and long GRBs. In contrast, the second component aligns closely with the long GRB population, exhibiting longer variability timescales. 
Similar trends are seen for \bat data in the 15–350 keV band, where we find that Episode 1 has an MVT of $0.50 \pm 0.17$\,s, while Episode 2 exhibits a significantly longer MVT of $27.37 \pm 0.15$\,s. These results suggest a notable transition in the temporal properties of \thisgrb between the two emission episodes.

\section{Afterglow Modelling}\label{sec:model}
\subsection{Model description}
We modelled the afterglow observations of GRB\,241105A within the standard synchrotron framework \citep{spn98}, where a point deposition of energy ($E_{\rm K,iso}$) into a circumstellar medium with density, $\rho=AR^{-k}$ generates a relativistic forward shock that accelerates electrons into a power-law distribution, $N_\gamma\propto\gamma^{-p}$ above a minimum Lorentz factor, $\gamma>\gamma_{\rm min}$.  We considered two standard possibilities for the radial density profile: a uniform density interstellar medium (ISM) model ($k=0$, $A\equiv n_0 m_p$; \citealt{spn98}) and a wind-like model ($k=2$, $A\equiv5\times10^{11}$g\,cm$^{-1}A_*$; \citealt{cl00}). The observed spectral shape of the resulting radiation is described by three break frequencies: the self-absorption frequency, $\nu_{\rm a}$; the characteristic injection frequency, $\nu_{\rm m}$; and the cooling frequency, $\nu_{\rm c}$; as well as the flux density normalization, $F_{\nu,\rm m}$ \citep{gs02}. We utilised the weighting prescription described in \cite{lbt+14} to compute smooth light curves during transitions between the different asymptotic regimes \citep{gs02}.  

We assumed that the afterglow jet is viewed on-axis, such that the above prescriptions (relevant for spherical outflows) apply. In addition, we utilised the formalism described in \cite{rho99} to incorporate the jet break effect. We accounted for inverse Compton (IC) cooling and Klein-Nishina corrections using the prescription provided by \cite{ml24}. We included Milky Way extinction of $A_{\rm V, MW}=0.03$~mag in our model \citep{sf11}, and assume a Small Megallanic Cloud (SMC)-type extinction for the host galaxy appropriate for GRB hosts \citep{smp+07,sdp+12}, with $A_{\rm V}$ as a free parameter \citep{pei92}. At radio wavelengths, we computed\footnote{We did not fold in the expected modulation due to scintillation as an additional source of systematic uncertainty for this dataset, as doing so significantly down-weights the radio data, which is a specific concern for this rather small dataset ($\sim27$ data points).} the effect of scintillation for visualization purposes, by computing the corresponding modulation index following the prescription of \cite{gn06} as described in \cite{lbt+14}. 

\subsection{Data used in afterglow modeling}
In addition to the UV/optical and radio observations reported here, we included X-ray data from \xrt, which we converted from a count-rate to a flux density at 1\,\keV\ using the spectral model reported on the UKSSDC website (photon index, $\Gamma_X=2.3$ and 0.3--10\,\keV\ unabsorbed counts-to-flux conversion factor, $3.9\times10^{-11}\,{\rm erg\,cm}^{-2}\,{\rm ct}^{-1}$). We further included two epochs of follow-up observations taken with the Follow-up X-ray Telescope on board \textit{Einstein Probe} at 0.92 and 1.58~days as reported by \cite{gcn38124}, which we also converted to flux density at 1\,\keV using the \xrt\ photon index. 

The wide filter bandpass of the \svom/VT R- and B- bands (henceforth, VT/R-band and VT/B-band, respectively), along with the Lyman-$\alpha$ absorption in the spectrum (which lies within the B-band) renders inter-band photometric comparison against other optical data challenging. To account for this, we derived a more accurate central wavelength for each filter given the observed spectrum. We fitted a continuum to the VLT spectrum using \texttt{contfit} in python and integrated the resultant model under the \svom/VT filters, from which we computed effective wavelengths of 5218\AA\ and 7750\AA\ for the VT/B- and R- bands, respectively. We note that the latter is actually closer to $i^{\prime}$-band. In addition, we found that the integrated flux within each band is lower (and therefore needs to be increased) relative to the values of the continuum at the effective wavelength by 19\% in the VT/B-band and 9\% in VT/R-band. The flux suppression is higher in VT/B-band due to Ly$\alpha$ absorption. Under the assumption that the spectral shape does not change significantly over the course of the \svom/VT observations, we applied these corrections uniformly to the VT data in subsequent discussion in this section. 

The NTT $r^\prime$ photometry at 1.47~days is a factor of 
2.5 fainter than the expected flux density of $\approx8\,\mu$Jy expected from interpolating between the VT/R-band points at $\approx0.8$ and $\approx2.2$~days. 
Additionally, the NTT $z^\prime$-$r^\prime$ spectral index\footnote{We use the convention, $F_\nu\propto t^\alpha\nu^\beta$ throughout.} of $\beta=0.5\pm0.7$ is \textit{positive} at this time, in strong contrast to the optical spectral index between the \svom/VT B and R-bands, which is $\beta_{R-B}=-2.08\pm0.16$ (even after the previously described correction, which increases the B-band flux more than that of the R-band). Similarly, the ATLAS $o$-band observation at 0.7~days, nominally in between the VT/B- and R-bands in center wavelength and obtained after the \svom\ observations at 0.67~days, is nevertheless brighter than both, and cannot be easily explained. We therefore did not include the NTT or ATLAS photometry in our analysis and instead plotted these data points on subsequent figures as open symbols. 

\subsection{Preliminary Modeling Considerations}
\label{text:prelim-optx}

\begin{figure*}
    \centering
    \begin{tabular}{cc}
         \includegraphics[width=\columnwidth]{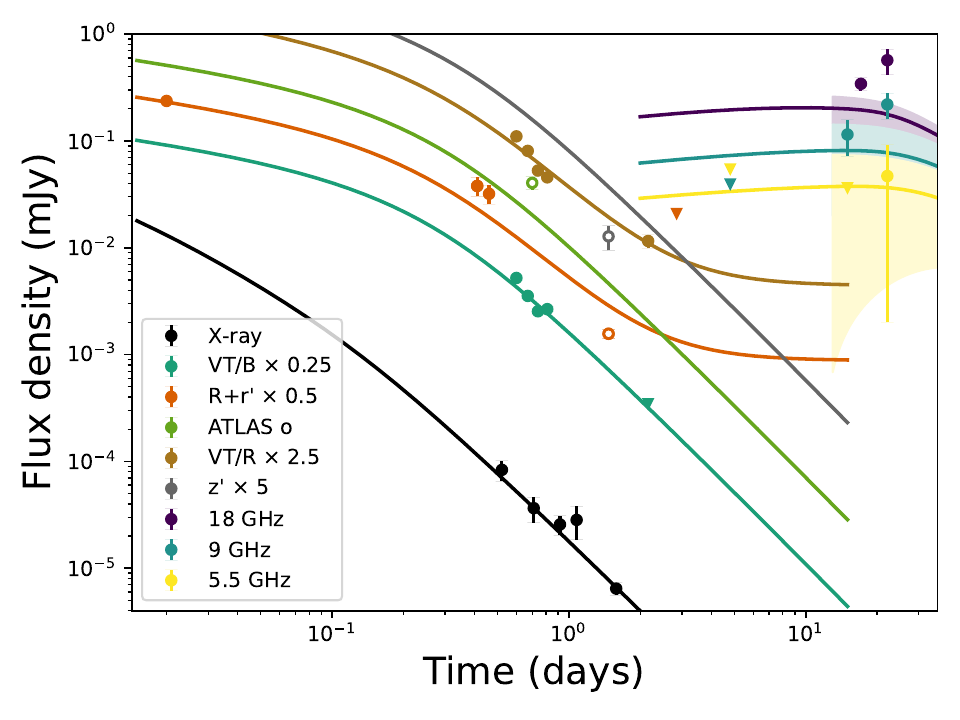} &  \includegraphics[width=\columnwidth]{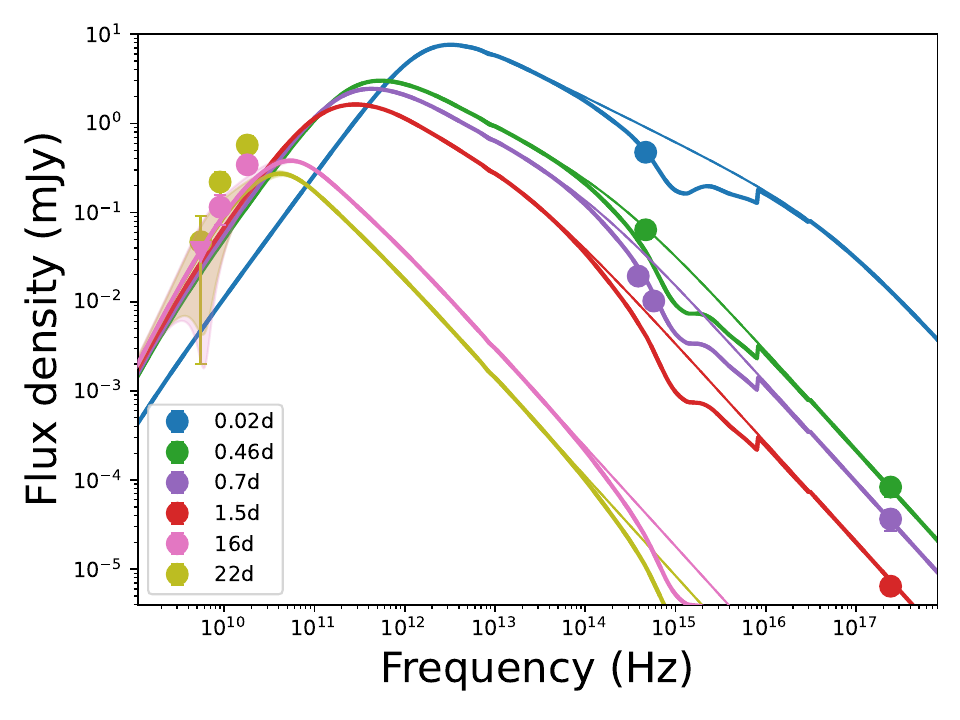}
    \end{tabular}    
    \caption{Light curves (left) and spectral energy distributions (SEDs, right) for our best-fit ISM model for the afterglow observations of GRB~241105A. Data points plotted in open symbols are not included in the analysis. Shaded bands indicate $1\sigma$ variability at each time expected from interstellar scintillation for reference. The flattening in the VT/R-band and $R+r^\prime$-band light curves is due to a fixed host contribution of $1.8\mu$Jy included in the modeling. Correlation contours for all physical parameters included in the fit, along with the derived parameters of  $\theta_{\rm jet}$ and $E_K$, are presented in Figure~\ref{fig:ag-corner-ISM}.}
    \label{fig:ag-ISM}
\end{figure*}
\begin{figure*}
    \centering
    \includegraphics[width=0.9\textwidth]{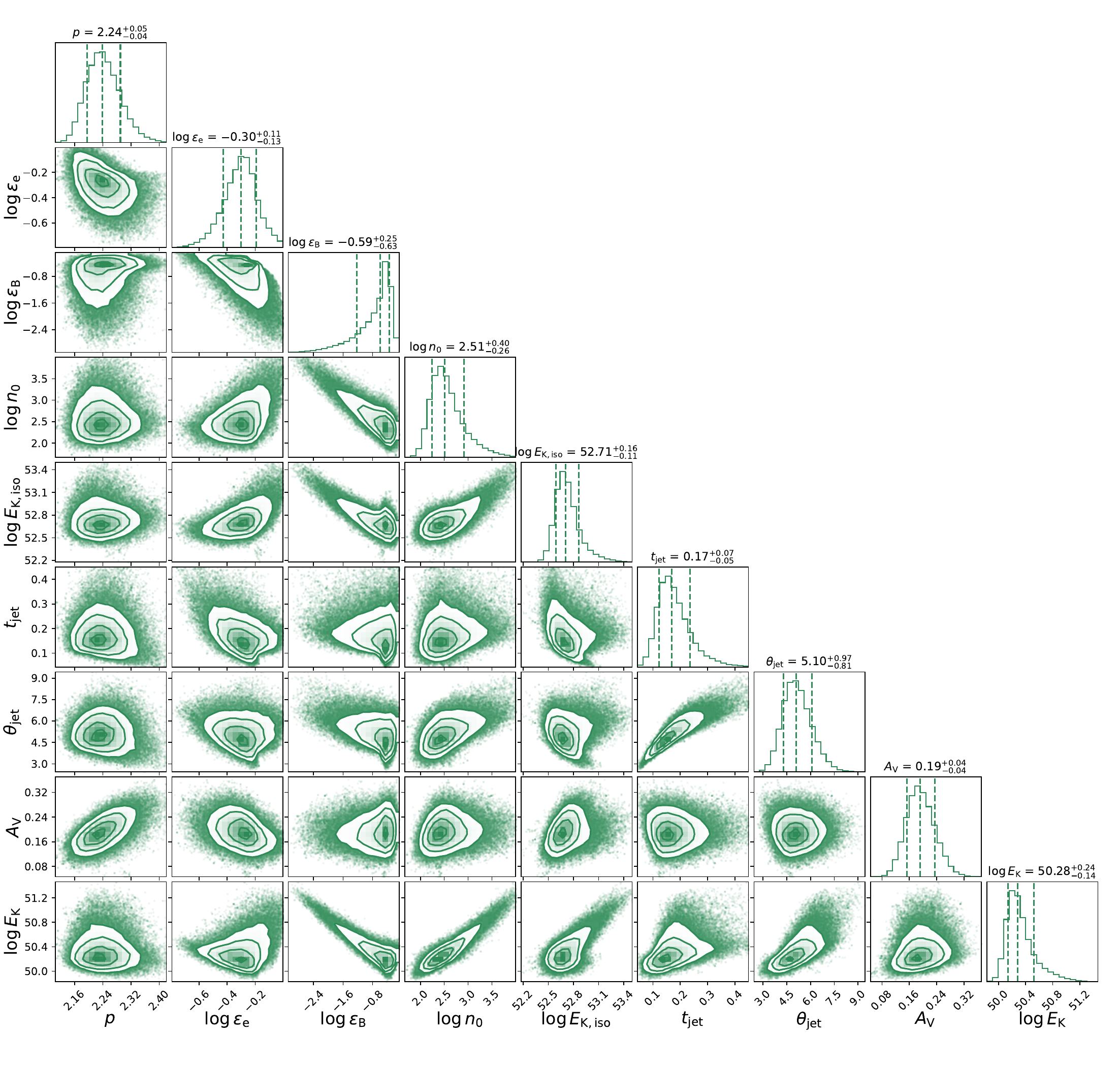}
    \vspace{-0.3in}
    \caption{Correlations and marginalised posterior density for all free parameters in the afterglow model. $E_{\rm K,iso}$ and $E_K$ are in units of erg, $t_{\rm jet}$ in days, $\theta_{\rm jet}$ in degrees, and $A_V$ in magnitudes. The contours enclose 39.3\%, 86.4\%, and 98.9\% of the probability mass in each correlation plot (corresponding to 1, 2, and 3$\sigma$ regions for two-dimensional Gaussian distributions, respectively. The dashed lines indicate the 15.9\%, 50\%, and 84.1\% quantiles, corresponding to the median and $\pm1\sigma$ for one-dimensional Gaussian distributions.}
    \label{fig:ag-corner-ISM}
\end{figure*}
\subsubsection{Optical and X-rays: Jet break, circumstellar density profile, host contribution, and relative locations of $\nu_{\rm m}$ and $\nu_{\rm c}$}
We further investigated the light curves (Fig.~\ref{fig:ag-ISM}, left panel) and spectral energy distributions (SEDs; Fig.~\ref{fig:ag-ISM}, right panel) of the afterglow to inform our modeling. The X-ray light curve comprising three \xrt\ points and two \textit{Einstein Probe}/FXT points can be fitted with a single power law with a steep decay, $\alpha_{\rm X}=-2.3\pm0.2$ from 0.5--1.6\,days. The optical light curve declines equally steeply during this period, with $\alpha_r^\prime=-1.9\pm0.3$ from the BOOTES-7 data point at 0.41~days and the Gemini-South observation at 1.4~days. The \svom\ optical light curves start at 0.6~days and are consistent with this steep decay, with $\alpha_{\rm VT-B}=-2.5\pm0.6$ and $\alpha_{\rm VT-R,1}=-3.0\pm0.6$. 
In the standard synchrotron framework, the steepest decline possible is in the regime $\nu_{\rm m}<\nu<\nu_{\rm c}$, with $\alpha=(1-3p)/4$ in the wind model. If the X-ray and optical bands are in this regime, then the X-ray decline rate would imply $p=3.4\pm0.3$ and require a spectral index of $\beta=(1-p)/2=-1.2\pm0.1$. The observed spectral index between the VT/R-band and the X-rays at $\approx0.67$~days is $\beta_{R-X}=-1.05\pm0.04$, which is consistent with the requirement. Although we are constrained in our modeling to $p<3$ by our KN correction framework, which does not allow us to consider values of $p>3$, we are able to find a model that matches the rough features of the light curves with $p\approx3$; however, this model under-predicts the radio observations by factors of 5--10, and we do not discuss this case further. 
Instead, a more natural explanation is afforded by considering the steep decline to be associated with post-jet break evolution. In the remainder of this discussion, we focused on that scenario, which then indicates $t_{\rm jet}\lesssim0.5$~days and $p\sim2.3$. 

Since the light curve evolution is the same in both ISM and wind cases after the jet break, and very little data exist prior to 0.5~days (Fig.~\ref{fig:ag-ISM}, left panel), it is not possible to determine the nature of the pre-explosion environment in this case. In the remainder of this section, we focused on the ISM case (which ends up providing a slightly lower reduced $\chi^2$, as we discuss in Section~\ref{text:mcmc}), and present the wind model in Appendix~\ref{appendix:wind}.  

The temporal decline between the last two VT/R-band points at 0.81 and 2.16~days (Fig.~\ref{fig:ag-ISM}, left panel) of $\alpha_{\rm VT-R,2}=-1.40\pm0.16$ is much shallower than $\alpha_{\rm VT-R,1}=-3.0\pm0.6$, consistent with the detection of the host galaxy in our JWST imaging (Section~\ref{obs:jwst}). In our subsequent analysis, we assumed a fixed\footnote{While it is possible to keep this parameter free, we found that doing so results in best-fit values differing by a factor of $\approx3$ between the two nearly identical bands, commensurate with the limited late-time ($\gtrsim1$~day) photometric information available. In this situation, fixing this parameter to a reasonable value yields better model convergence.} (and equal) host contribution in the $r^\prime$ and VT/R bands corresponding to a flux density of $1.8\,\mu$Jy of the host measured in a 1$\arcsec$ aperture within the closest JWST filter available (F070W; Table~\ref{tab:jwst_photometry}). 

The temporal decay of $\alpha=-0.64\pm0.06$ between the GOTO $L$-band observation at 0.02~days and the VLT $r^\prime$-band observation at 0.46~days is shallower than the slowest afterglow decay rate possible, which is $\alpha=(3-3p)/4=-0.98\pm0.15$ in the $\nu_{\rm m}<\nu<\nu_{\rm c}$ regime. Instead, the fast-cooling model $\nu_{\rm c}<\nu<\nu_{\rm m}$ allows for the additional possibility of $\alpha=-2/3$ in the $\nu_{\rm a}<\nu<\nu_{\rm c}$ regime (wind case) and $\alpha=-1/4$ in the $\nu_{\rm c}<\nu<\nu_{\rm m}$ regime (wind or ISM case). We found that both are plausible; the first matches the observed decay closely, while the latter can still be accommodated with moving the jet break earlier. In either case, a fast cooling segment in the early light curves ($\lesssim0.5$~days) is required, indicative of either a high magnetic field or a high-density environment, both of which can result in a lower cooling frequency. We return to this point in section~\ref{text:prelim-radio}. 

The spectral index between the VT/R-band and VT/B-band, $\beta_{\rm R-B}=-2.08\pm0.16$, is much steeper than the spectral index from either the VT/R-band or the VT/B-band to the X-rays, which are $\beta_{R-X}=-1.05\pm0.04$ and $\beta_{B-X}=-0.99\pm0.05$, respectively, indicating that extinction is present\footnote{The X-ray spectral index from \swift/XRT, $\beta_{\rm X}=-1.3^{+0.8}_{-0.7}$ is consistent with the optical-to-X-ray spectral index, but is otherwise not constraining due to the large uncertainty.} (Fig.~\ref{fig:ag-ISM}, right panel). The observed spectral index of $\beta\approx-1$ between the optical and X-rays is marginally shallower than the spectral index expected in the regime ${\nu_{\rm m},\nu_{\rm c}<\nu}$ of $\beta=-p/2=-1.15\pm0.10$, which may arise from either the aforementioned extinction or from a spectral break proximate to the optical wavelengths. Since the afterglow is expected to be in fast cooling, this is suggestive of the presence of $\nu_{\rm c}$ close to the optical bands. We return to this point in section~\ref{text:mcmc}.

\subsubsection{Radio: synchrotron self-absorption}
\label{text:prelim-radio}
A power-law fit to the radio detections at $\approx22$~days yields a spectral index of $\beta_{\rm radio,22}=1.7\pm0.4$, which is consistent with the spectral index between the ATCA X-band (9.0\,GHz) and Ku/15mm band (18\,GHz) at 15--17~days $\beta_{\rm XU}=2.4\pm1.0$ as well as with the 5~GHz non-detection in this earlier epoch, together indicating a strongly self-absorbed synchrotron spectrum, $\nu_{\rm a}\gtrsim18$\,GHz until $\approx22$~days, indicative of a high-density pre-explosion environment. This is consistent with the low cooling frequency inferred from the early ($\lesssim0.5$~day) optical light curve, as discussed in section~\ref{text:prelim-optx}. 

In the framework where the radio emission arises from the afterglow forward shock and in the scenario that the jet break ($t_{\rm jet}\lesssim0.5$~days) has already occurred at by the time the radio observations commence, the radio light curves are expected to be constant in this regime; instead, the light curves appear to rise both at 18~GHz ($\alpha_{\rm 18\,GHz}=2.0\pm1.3$) and at 9~GHz ($\alpha_{\rm 9\,GHz}=1.7\pm1.3$; and also at 5.5~GHz, although the rise rate in that band is much harder to measure, since the first two epochs yielded non-detections and the sole detection in the third epoch has very large uncertainties). This behaviour is unexpected in the standard framework (although, see also \citealt{vmwk11}). One possibility is that this may be due to scintillation; however, detailed analysis of this possibility requires additional radio observations. We nevertheless included the radio observations in our modeling and discuss deviations from the best-fit model further in section~\ref{text:mcmc}. 

\begin{table}
    \caption{Afterglow Model Parameters}
    \begin{threeparttable}
    \centering
    \begin{tabular}{lrrrr}
        \toprule
        & \multicolumn{2}{c}{ISM} & \multicolumn{2}{c}{Wind$^{\rm a}$} \\
        \cmidrule(lr){2-3} \cmidrule(lr){4-5}
        Parameter & Best fit & MCMC & Best fit & MCMC \\
        \midrule
$p$      & $2.24$ & $2.24^{+0.05}_{-0.04}$  & $2.21$ & $2.21^{+0.05}_{-0.04}$\\[2pt]
$\log\epsilon_{\rm e}$  & $-0.31$ & $-0.30^{+0.11}_{-0.13}$ & $-0.31$ & $-0.23^{+0.14}_{-0.17}$ \\[2pt]
$\log\epsilon_{\rm B}$  & $-0.37$ & $-0.59^{+0.25}_{-0.63}$ & $-0.81$ & $-1.21^{+0.67}_{-1.01}$ \\[2pt]
$\log n_0$ & $2.27$ & $2.51^{+0.40}_{-0.26}$ & \ldots & \ldots \\[2pt]
$\log A_*$ & \ldots & \ldots & $0.40$ & $0.63^{+0.55}_{-0.32}$ \\[2pt]
$\log(E_{\rm K,iso}$/erg) & $52.7$ & $52.71^{+0.16}_{-0.11}$ & $52.7$ & $52.75^{+0.35}_{-0.19}$ \\[2pt]
$A_V/$mag & $0.19$ & $0.19\pm0.04$ & $0.20$ & $0.19\pm0.04$ \\[2pt]
$t_{\rm jet}$/days & $0.13$ & $0.17^{+0.07}_{-0.05}$ & $0.17$ & $0.20^{+0.08}_{-0.05}$ \\[2pt]
$\theta_{\rm jet}$/deg & $4.31$ & $5.10^{+0.97}_{-0.81}$ & $4.62$ & $5.23^{+1.11}_{-0.85}$ \\[2pt]
$\log(E_{\rm K}$/erg) & $50.1$ & $50.28^{+0.24}_{-0.14}$ & $50.2$ & $50.35^{+0.47}_{-0.25}$ \\[2pt]
$\log(\nu_{\rm ac}$/Hz)$^{\rm b}$ & $9.1$  & \ldots & $9.2$  & \ldots\\[2pt]
$\log(\nu_{\rm sa}$/Hz) & $11.9$ & \ldots & $12.0$ & \ldots\\[2pt]
$\log(\nu_{\rm c}$/Hz)$^{\rm b}$  & $10.5$ & \ldots & $9.0$  & \ldots\\[2pt]
$\log(\nu_{\rm m}$/Hz)  & $15.6$ & \ldots & $15.4$ & \ldots\\[2pt]
$F_{\nu,\rm sa}$/mJy    & $15.9$ & \ldots & $13.6$ & \ldots\\[2pt]
\bottomrule
    \end{tabular}
    \label{tab:afterglow-parameters}
    \begin{tablenotes}
        \item[a] The wind model is discussed further in Appendix~\ref{appendix:wind}.
        \item[b] This break frequency is not directly constrained by the data.
        \item All break frequencies and fluxes are provided at $0.1$~days. 
    \end{tablenotes}
    \end{threeparttable}
\end{table}

\subsection{Markov Chain Monte Carlo Analysis}
\label{text:mcmc}
We explored the multi-dimensional likelihood space for the parameters of the afterglow model using a Markov Chain Monte Carlo with \texttt{emcee}. Further details
of the modeling procedure, including the likelihood function employed, are described elsewhere \citep{lbz+13,lbt+14,lab+16}. To account for flux calibration offsets between observations across telescopes, we included a systematic uncertainty of 10\% added in quadrature to all measurements. We employed uniform priors on $p$ and the extinction, $A_V$ and \cite{jef46} priors for the remaining parameters ($\epsilon_{\rm e}$, $\epsilon_{\rm B}$, $E_{\rm K,iso,52}$, $n_0$, $A_*$, and $t_{\rm jet}$). We run 512 chains for 2000 iterations (testing for convergence by tracking the mean likelihood with iteration) and discard the initial $50$ steps as burn-in.

As discussed above, we focused on the ISM model here and present the wind model in Appendix~\ref{appendix:wind}. Our best-fit ISM model has $p=2.24$, $\epsilon_{\rm e} = 0.49$, $\epsilon_{\rm B} = 0.43$, $n_0 = 1.9\times10^2$, $E_{\rm K,iso}=4.8\times10^{52}$~erg, $t_{\rm jet}=0.13$~days, and $A_V=0.19$~mag. 
For these parameters, $\nu_{\rm m}\approx4.1\times10^{15}$~Hz at $\approx0.1$~days. This break passes through the optical bands between 0.1--0.4~days, consistent with the expectations in Section~\ref{text:prelim-optx}. The afterglow is in fast cooling with $\nu_{\rm c}<\nu_{\rm m}$. In this regime, the self-absorption break splits into two ($\nu_{\rm ac}$ and $\nu_{\rm sa}$), and for these parameters the spectral peak is at $\nu_{\rm sa}\approx7.6\times10^{11}$~Hz with a peak flux density of $\approx16$~mJy. The cooling break, $\nu_{\rm c}\approx3\times10^{10}$~Hz at this time and is unobservable, while $\nu_{\rm ac}\approx1.1\times10^{9}$~Hz only crosses above 5.5~GHz at $\approx7.5$~days and is, therefore, weakly constrained.
In this spectral regime\footnote{We note that in this regime where $\nu_{\rm sa},\nu_{\rm c}<\nu<\nu_{\rm m}$ corresponding to the $F_\nu\propto\nu^{-1/2}$ spectral power-law segment, the optical light curves are insensitive to the density profile.}, the decline rate in the optical bands is $\alpha=-1/4$; this is steepened in the model light curves by the jet break, and the optical light curves further steepen subsequently when $\nu_{\rm m}$ passes through the observing band. 
We plot light curves and SEDs from the radio to X-rays of the best-fit model in Figure~\ref{fig:ag-ISM} and provide corner plots of the correlations and marginalised posterior density for all fitted parameters (and the derived parameters, $\theta_{\rm jet}$ and $E_K$) from our MCMC analysis in Figure~\ref{fig:ag-corner-ISM}. We list our derived parameters in Table~\ref{tab:afterglow-parameters}.

The contour plots reveal correlations between the physical parameters. Upon investigation, we associate these degeneracies with a range of allowed values for the Compton Y-parameter that nevertheless result in similar light curves and SEDs in the absence of a clear identification of the cooling break (the latter being hidden below $\nu_{\rm sa}$ for all parameter ranges spanned by this model). Following the framework of \cite{lbt+14}, we quantify these degeneracies by explicitly asking which segments of the SEDs are constrained by the data. From Fig.~\ref{fig:ag-ISM}, we identify these to be the flux on power law segments C, F, and H \citep{gs02}, $F_{\rm C} \equiv F_{\nu_{\rm ac}<\nu<\nu_{\rm sa}}$, $F_{\rm F} \equiv F_{\nu_{\rm sa}<\nu<\nu_{\rm m}}$, and $F_{\rm H} \equiv F_{\nu<\nu_{\rm m}}$, which are constrained by the radio, optical, and X-ray observations, respectively. Solving for the physical parameters in terms of these three fluxes and the (unobserved) cooling frequency, while  accounting for the expected effects of IC cooling as described in \cite{gs02}, we expect 
$\epsilon_{\rm e}\propto\nu_{\rm c}^{-1/4}$,
$\epsilon_{\rm B}\propto\nu_{\rm c}^{7/4}$, 
$n_0\propto\nu_{\rm c}^{-5/4}$, and 
$E_{\rm K,iso}\propto\nu_{\rm c}^{-3/4}$. The resulting expected correlations between the physical parameters due to the unknown value of $\nu_{\rm c}$ are then $E_{\rm K,iso}\propto n_0^{3/5}\propto \epsilon_{\rm B}^{-3/7}\propto \epsilon_{\rm e}^{3}$, $n_0\propto\epsilon_{\rm e}^{5}\propto\epsilon_{\rm B}^{-5/7}$, and $\epsilon_{\rm B}\propto\epsilon_{\rm e}^{-7}$, all of which are fully consistent with the sense and roughly consistent with the slopes of the corresponding observed correlations. Since $\nu_{\rm c}$ is unobservable in this model\footnote{We note that in the standard framework ignoring IC cooling, $\nu_{\rm c}$ can be inferred using a combination of $\nu_{\rm sa}$, $\nu_{\rm ac}$, and $\nu_{\rm m}$; however, the addition of Compton $Y$ as an additional parameter results in the specific degeneracies described here.}, few observational constraints could help resolve this particular degeneracy.

\section{Host Characteristics}
\label{sec:host}
\subsection{Kinematics of the Host Environment}
As described in Section~\ref{sec:observations}, the VLT/FORS2 spectrum of the afterglow of \thisgrb shows a strong Ly$\alpha$ absorption trough and numerous metal absorption lines at a common redshift of $z = 2.681$, which we adopt as the systemic redshift of the host galaxy.
We estimated the neutral hydrogen column density ($N_{\rm HI}$) along the line of sight to GRB~241105A from the broad Ly$\alpha$ absorption feature observed at the host redshift. The absorption profile is modeled using {\tt VoigtFit} \citep{Krogager2018}, incorporating the spectral resolution of the data. The redshift ($z$) and Doppler parameter ($b$) were fixed to the values derived from metal line fitting (see below), leaving $N_{\rm HI}$ as the only free parameter. The best-fit model yields $\log(N_{\rm HI}/\mathrm{cm}^{-2}) = 21.15 \pm 0.10$, consistent with values commonly observed in long GRB hosts at $z \gtrsim 2$ \citep{Jakobsson2006,Fynbo2009,Selsing2019,Heintz2023}.
\begin{table*}
\centering
\caption{List of transitions and corresponding column densities of low- and high-ionization absorption lines identified in the GRB\,241105A VLT/FORS2 spectrum using the Voigt fit procedure. All listed transitions are observed at the common redshift $z=2.681$. Column density measurements are derived assuming a Doppler parameter $b=118 \pm 20$\,km\,s$^{-1}$ from the simultaneous fit of low-ionization transitions with a single component. High-ionization transitions exhibit a blueward shift of $\sim -150$\,km\,s$^{-1}$ and a larger Doppler parameter $b=143 \pm 20$\,km\,s$^{-1}$. Saturated transitions are reported as 3$\sigma$ lower limits to maintain conservatism given resolution and hidden saturation effects.}
\label{column}
\renewcommand{\arraystretch}{1.3}
\begin{threeparttable}
\begin{tabular}{lc}
\toprule
Transitions (\AA) & $\log (N/\mathrm{cm}^{-2})$ \\
\midrule
\sii{} $\lambda1253, \lambda1255, \lambda1259$                          & $>15.0$ \\
\siii{} $\lambda1260, \lambda1304, \lambda1527, \lambda1808$            & $16.1 \pm 0.2$ \\
\siii{}{*} $\lambda1264, \lambda1265, \lambda1533$                     & $14.0 \pm 0.2$ \\
\oi{} $\lambda1302$                                                    & $>16.5$ \\
\cii{} $\lambda1334$                                                   & $>16.0$ \\
\cii{}{*} $\lambda1335$                                                & $>14.5$ \\
\feii{} $\lambda1608, \lambda2344, \lambda2374, \lambda2382, \lambda2587, \lambda2600$ & $15.1 \pm 0.2$ \\
\alii{} $\lambda1670$                                                  & $>14.2$ \\
\midrule
\nv{} $\lambda1238, \lambda1242$                                      & $>14.4$ \\
\siiv{} $\lambda1393, \lambda1402$                                    & $>15.1$ \\
\civ{} $\lambda1548, \lambda1550$                                     & $>14.4$ \\
\aliii{} $\lambda1854, \lambda1862$                                   & $>13.9$ \\
\bottomrule
\end{tabular}
\end{threeparttable}
\end{table*}

\begin{figure}
        \centering
    \includegraphics[width=\columnwidth]{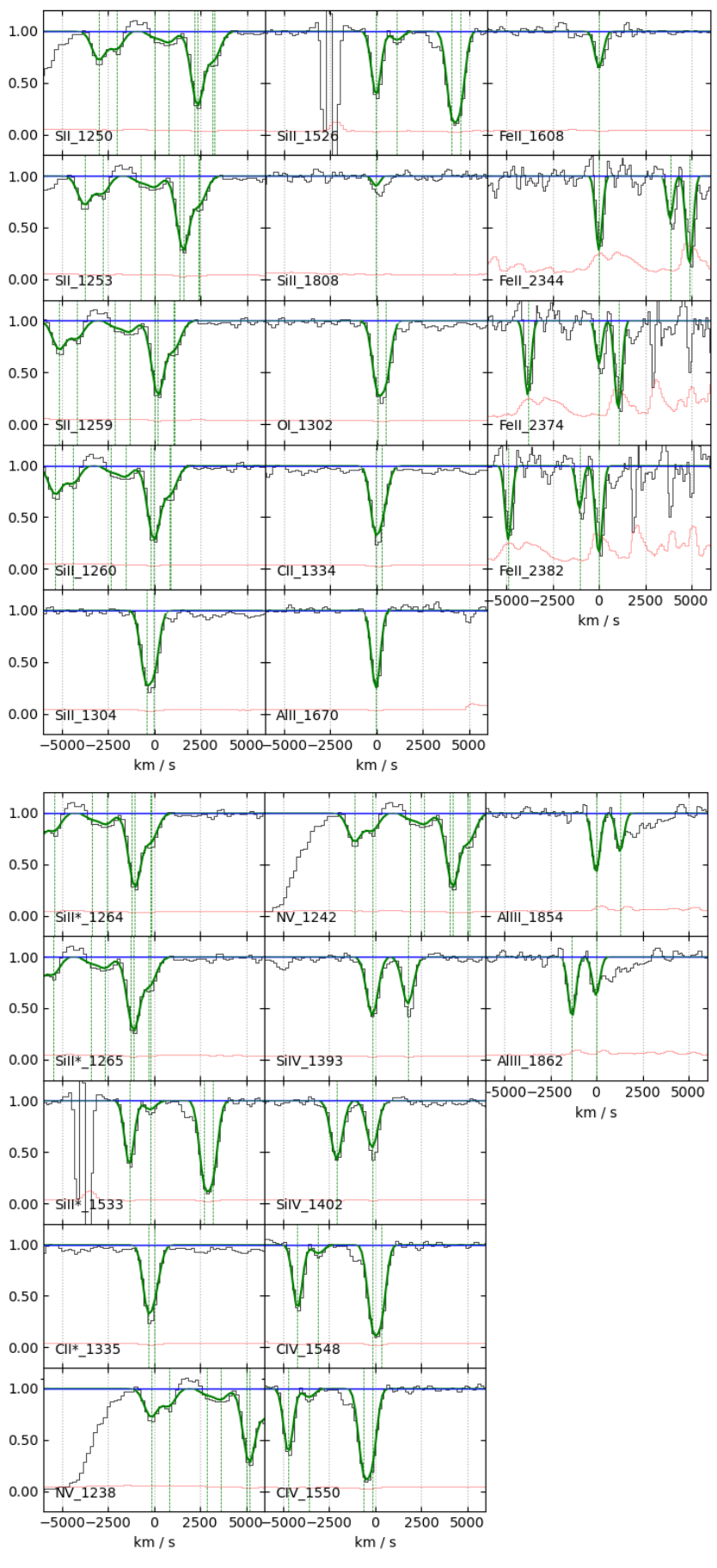}
    \caption{VLT/FORS2 optical afterglow spectrum of GRB\,241105A at redshift $z=2.681$. {\it Top panel}: low-ionization absorption lines of the GRB host galaxy system. Here and in the bottom panel data are in black, the fit is in green, the error spectrum is in red, the continuum in blue and the vertical green dashed lines indicate the center of the components. {\it Bottom panel}: fine structure lines and high-ionization absorption lines.}
    \label{abs_spec}
\end{figure}

We fitted the identified metal absorption lines with the Astrocook software \citep{Cupani2020}, a Python code environment to analyze spectra modelled with Voigt profiles, depending on the system redshift $z$, its column density $N$, and its Doppler broadening {\it b}. The Voigt profile fitting of each identified transition is shown in Figure \ref{abs_spec} and the corresponding column densities are reported in Table~\ref{column}. We infer a high Doppler parameter (i.e. $b=118\pm20$\,km~s$^{-1}$ for low-ionization absorption lines)  given the resolution of the spectrum, hence we are not able to really dissect the ISM into several gas cloud components. Furthermore, several absorption lines are saturated and we decide to provide 3$\sigma$ column density lower limits. There are a few exceptions such as \siii{}, \siii{*}{} and \feii{}, given the flux residuals of some transitions belonging to the same multiplet which allowed us to provide a column density measurement. 

To cross-validate the Voigt profile fitting, we also performed a curve-of-growth (CoG) analysis using a subset of unblended lines of low-ionization electronic transitions. The Doppler parameter obtained by constraining the CoG with several equivalent width (EW) measurements (by using a combination
of the lines with different oscillator strengths) is $b = 112.3 \pm 35.8$\,km\,s$^{-1}$, consistent with the Voigt results. A potential overestimate in the column density of \ion{C}{II} is noted due to unresolved blending with \ion{C}{II*}. Further details of the CoG fitting procedure and results are provided in the Appendix~\ref{sec:CoG}.
\begin{figure*}
    \centering
    \includegraphics[width=0.99\columnwidth]{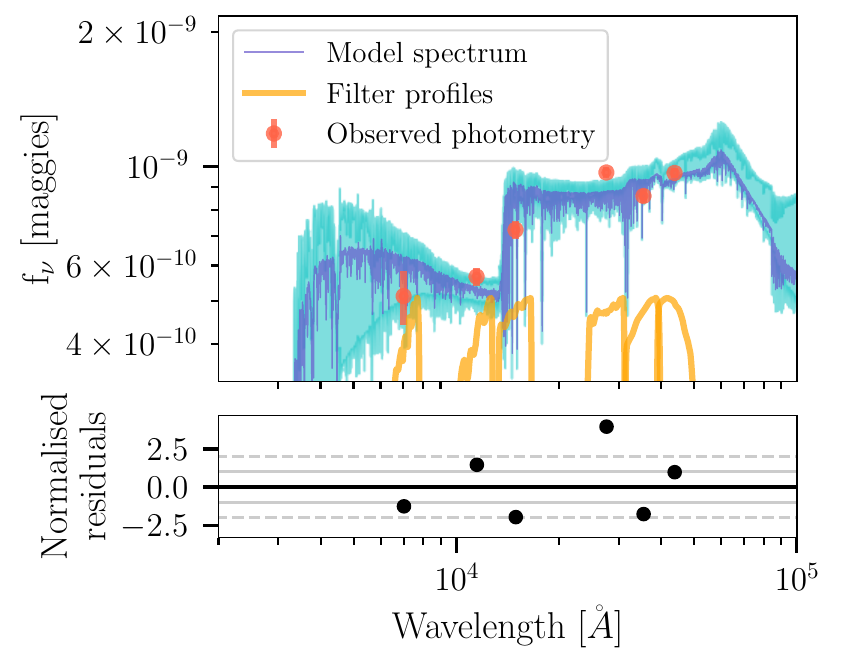}
    \includegraphics[width=0.95\columnwidth]{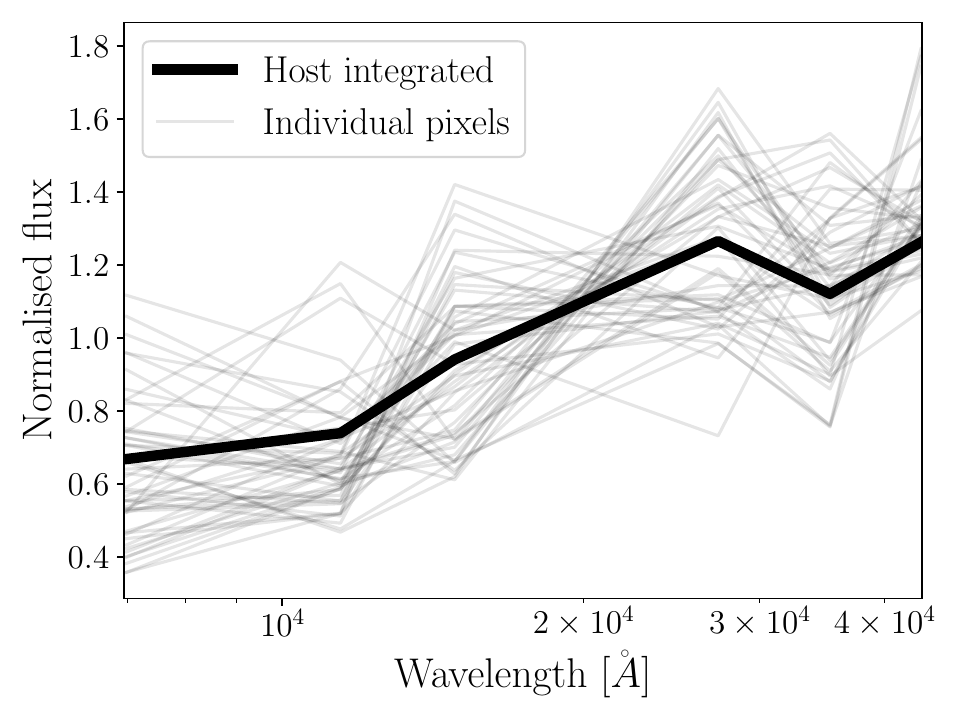}
    \caption{\textbf{\emph Left:} Results of the \textsc{Prospector} fit to the host photometry. \textit{Upper panel:} Best fitting Spectral Energy Distribution (SED) in blue, the shaded light blue region represents the 2\,$\sigma$ uncertainty of the predicted SED at each wavelength. All wavelengths are in the observer frame. The orange markers are the observations with the error bars representing uncertainties on the vertical axis, and JWST filter profiles indicated in yellow on the horizontal axis. \textit{Lower panel:} Normalised residuals between the observed photometry and the best fit model.
    \textbf{\emph Right:} Spectral energy distributions across the six JWST filters, normalised by the mean flux, for every pixel in the 10$\times$10\,pix region around the transient location (light grey lines). The solid black line is the host-integrated SED, used for SED fitting as described in Section~\ref{sec:sedfitting}. The overall SED shape is broadly similar across the host.
    }
    \label{fig:Prospector_SED_pixelmap}
\end{figure*}
\begin{figure*}
    \centering
    \includegraphics[width=\textwidth]{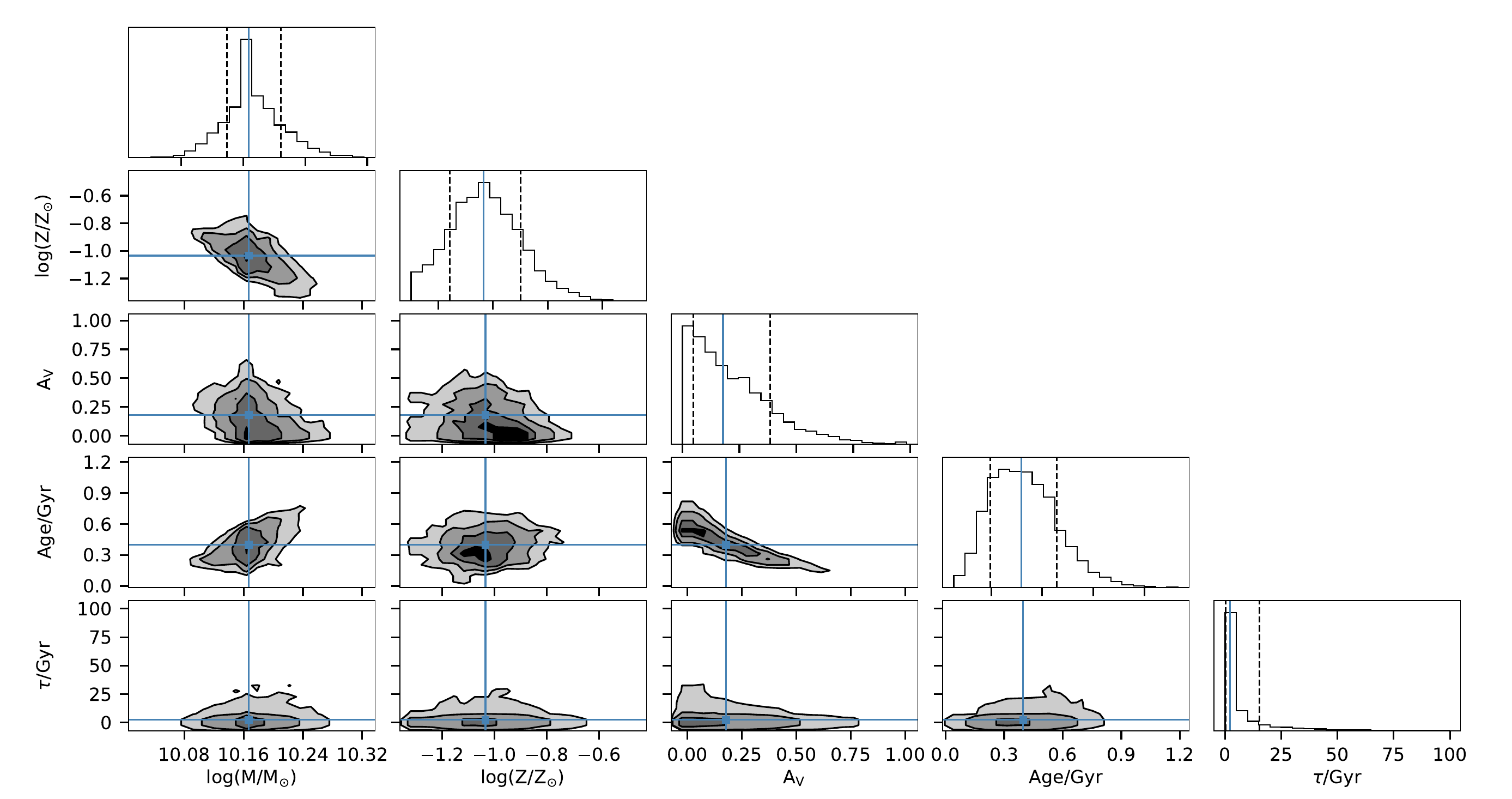}
    \caption{Corner plot for the posterior distribution of the parameters used in the SED fit for the host galaxy of GRB 241105A. This figure was created using the \textsc{corner.py} package \citep{ForemanMackey2016}. The best-fit parameters and their uncertainties are provided in Table~\ref{tab:host_props}.}
    \label{fig:Prospector_SED_triangle}
\end{figure*}

\begin{table}
  \centering
  \caption{Host-galaxy properties of \thisgrb. The uncertainties correspond to 1$\sigma$ confidence intervals.}
  \renewcommand{\arraystretch}{1.4}
  \begin{threeparttable}
    \begin{tabular}{lc}
      \hline
      Parameter & Value \\
      \hline
      $\log(M_*/M_\odot)$ & $10.16^{+0.04}_{-0.02}$ \\
      SFR [M$_\odot$\,yr$^{-1}$] & $135^{+456}_{-102}$ \\
      $t_{\rm age}$ (Gyr) & $0.37^{+0.20}_{-0.14}$ \\
      $\log(Z/Z_\odot)$ & $-1.03^{+0.13}_{-0.12}$ \\
      $A_V$ (mag) & $0.18^{+0.25}_{-0.14}$ \\
      Offset (kpc) & $0.18^{+0.29}_{-0.18}$ \\
      \hline
    \end{tabular}
  \end{threeparttable}
  \label{tab:host_props}
\end{table}

\subsection{Host SED fitting}\label{sec:sedfitting}
We fit the available JWST photometry with \textsc{Prospector} \citep{Johnson2021}, with a delay $\tau$ Star Formation History (SFH) and {\sc bpass} for the spectral synthesis \citep{2017PASA...34...58E,2018MNRAS.479...75S}. 
We let the stellar mass, visual attenuation, SFH e-folding time, and burst age vary freely within flat priors, and explored the parameter space using Markov chain Monte Carlo sampling with {\sc emcee} \citep{2013PASP..125..306F}. Metallicity is treated with a top hat prior, with lower and upper bounds of log$_{10}$(Z/Z$_{\odot}$ = -1.3 and 0.3 respectively. The lower bound is constrained by [S/H], as measured in the afterglow spectrum. The predicted SED of the host galaxy, along with the measured photometry, is shown in the left panel of Figure~\ref{fig:Prospector_SED_pixelmap}.
The properties derived from the fit shown in Table~\ref{tab:host_props} (see also Figure~\ref{fig:Prospector_SED_triangle}) indicate the host galaxy is rather massive, star-forming, and with low-metallicity.

To investigate whether there are significant variations across the host, we aligned the six JWST images, degrading the short-wavelength filters to the same pixel scale as the long-wavelengths, and smoothing each with a Gaussian filter to match the FWHM of the F070W image. We then selected pixels with a flux level more than 5$\sigma$ above the background in all six filters, in a 10$\times$10\,pixel region around the GRB location, and plotted their SEDs in the right panel of Figure \ref{fig:Prospector_SED_pixelmap}. Although the pixel SEDs show deviations on a filter to filter level, the overall SED shape is similar across the host. We therefore deem the host-integrated properties (namely extinction, age, and metallicity) to be representative of the transient location, although we caution that metallicities from SED fitting do suffer from systematics \citep{Leja2019, Johnson2021}. 

\subsection{GRB offset and host morphology}\label{sec:offset}
We determined the position of the afterglow in the JWST reference frame by identifying sources in common between the stacked FORS2 $g$-band detection images, and the F070W JWST image. Using DAO star finder routines in \texttt{Photutils} module of  \texttt{Python} \citep{larry_bradley_2024}, we found 5 good cross-match objects which are compact and symmetrical. The field is relatively sparse with few suitable cross-match objects, others were available but rejected due to saturation or diffuse/extended emission (i.e. galaxies) for which the centroid is less certain. We fixed the transformation between the two known pixel scales and allow for x-y shifts and rotation, to find a mapping from the FORS2 reference frame to that of the JWST image which minimises the root-mean-square (rms) of the positional offsets post-transformation. The rms of the final transformation is 36\,mas. The positional uncertainty of the afterglow centroid in the FORS2 image (given by the FWHM/(2.35$\times$SNR)) is 16\,mas, for a total positional uncertainty of the afterglow in the JWST frame of 39\,mas. The burst location is shown in Figure \ref{fig:location}.

\begin{figure}
     \centering
     \includegraphics[width=\columnwidth]{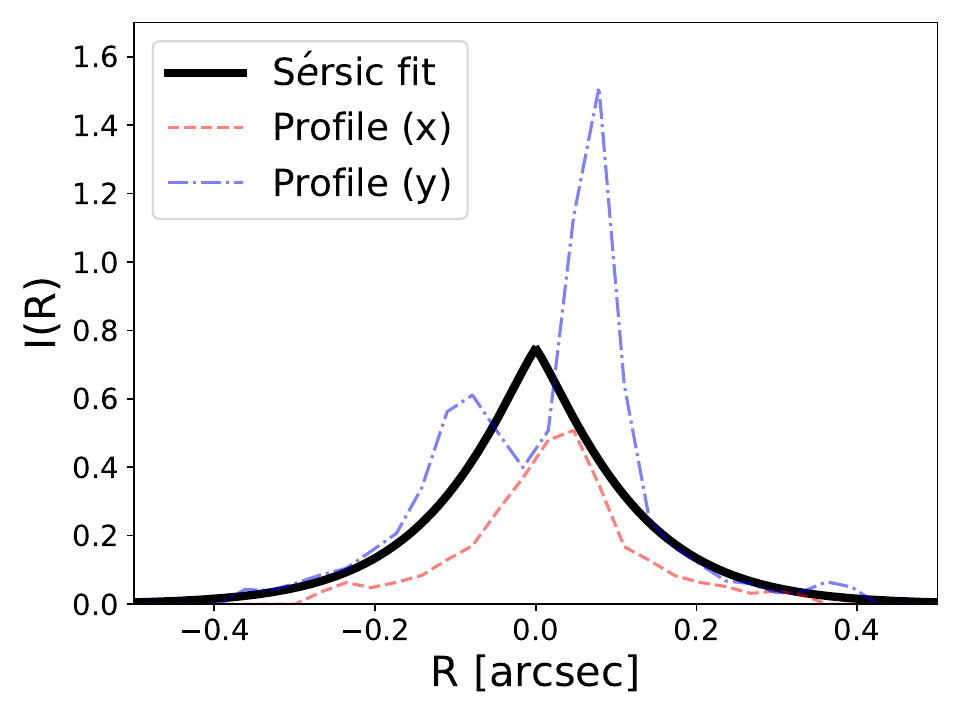}
     \caption{The S\'ersic profile fit to the host galaxy, as described in Section \ref{sec:offset}. 1D slices through the host centroid in $x$ and $y$ are also shown.}
     \label{fig:sersic}
\end{figure}
We also measured the morphological properties of the host galaxy with {\sc statmorph} \citep{Rodriguez-Gomez_2019}. A S\'ersic fit (shown in Figure \ref{fig:sersic} yields a half-light radius $r_{50} = 170$\,mas, if we take the galaxy (which is composed of two prominent clumps) as one object. At $z=2.681$ this corresponds to a physical size of 1.4\,kpc. Using the x,y centroid of the galaxy from the {\sc statmorph} S\'ersic fit and the refined afterglow position, we found a projected host offset of 23$^{+37}_{-23}$\,mas, or 0.18$^{+0.29}_{-0.18}$\,kpc. The uncertainties are at the 1$\sigma$ level and are calculated assuming a Rice distribution. The corresponding host-normalised offset is $r_{\rm norm} = 0.13^{+0.21}_{-0.13}$. Such a low projected offset is routinely observed for collapsar GRBs (albeit at the low end of the distribution), and extremely rare for compact object merger progenitors \citep[e.g.][]{Fong_2022}. Furthermore, the concentration and asymmetry values for the host galaxy ($C=2.97$ and $A=0.27$, respectively) lie in the spiral/irregular/merger region in C-A parameter space. This is consistent with its star-forming nature, and with the population of collapsar host galaxies \citep{Conselice_2005, Schneider2022}.

\section{Discussions}
\label{sec:discussion}
\subsection{\textbf{An Ambiguous "short+EE" Burst}}
\thisgrb, with a initial short spike followed by prolonged, weaker emission, places it firmly within the so-called "short+EE" GRB population. While this two-component profile is a defining characteristic of these events, \thisgrb exhibits a crucial difference that sets it apart from the classical examples of merger-driven EE GRBs like GRB 060614, GRB 211211A, and GRB 230307A. The EE in the case of \thisgrb is not as spectrally soft as typically observed in these well-studied events. 
Similar behavior has been noted in other nominally short GRBs at higher redshifts \citep{Dichiara_2021}, where EE is seen in the higher energy bands for short GRBs with $z>1$, challenging the conventional definition of EE GRBs.
\begin{figure*}
    \centering  \includegraphics[height= 4.5cm, width=0.33\linewidth]{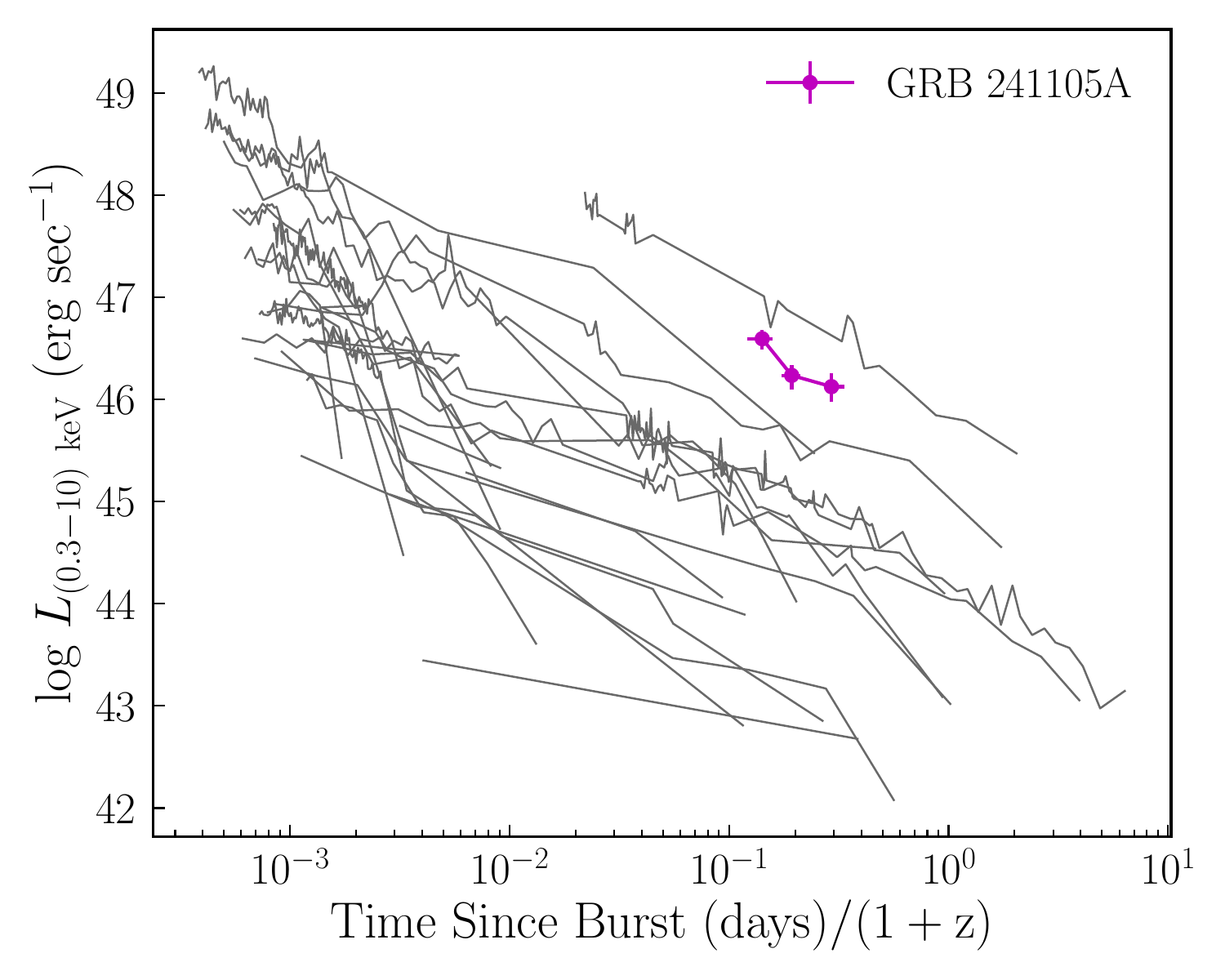}
    \includegraphics[height= 4.5cm, width=0.33\linewidth]{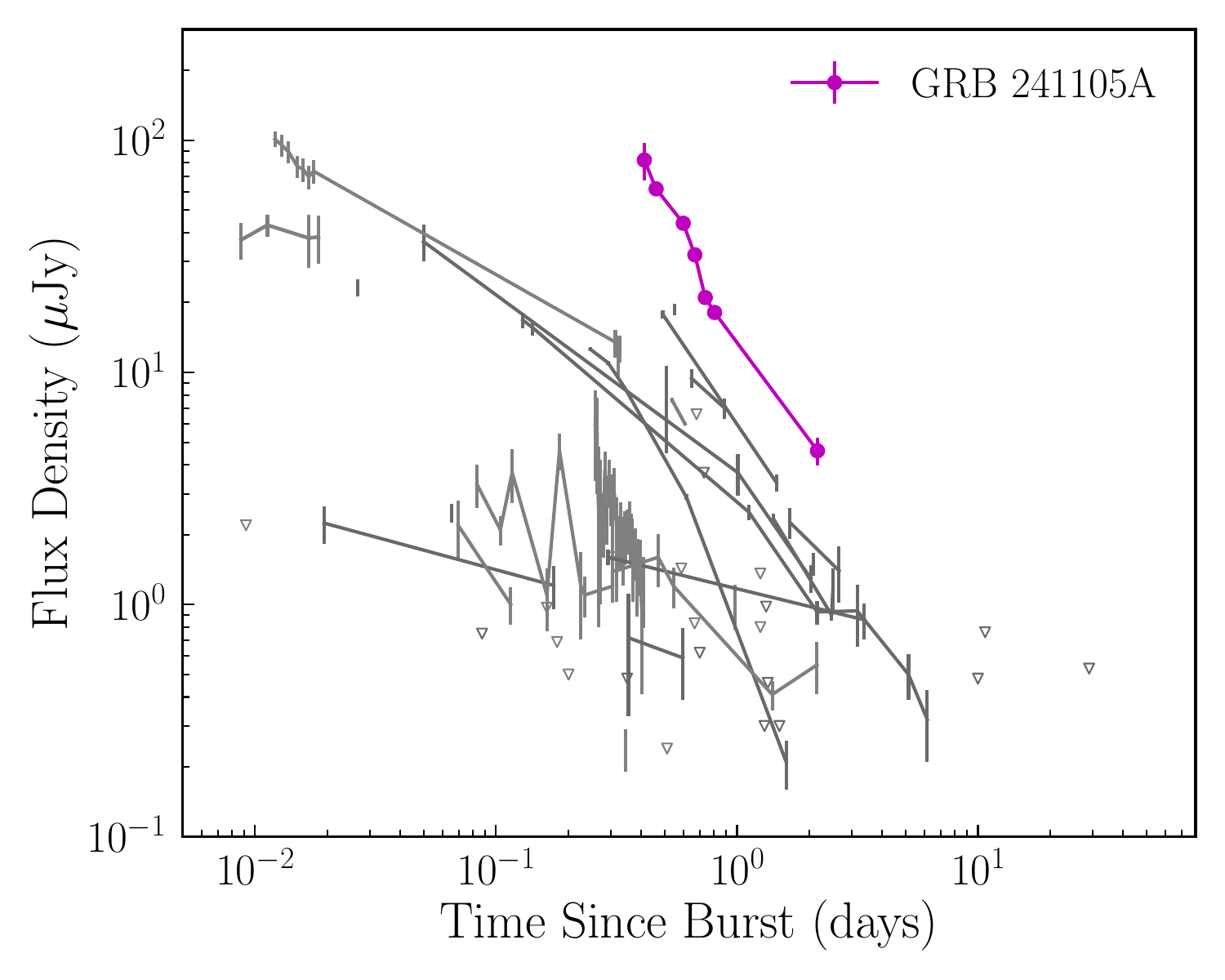}
    \includegraphics[height= 4.5cm, width=0.33\linewidth]{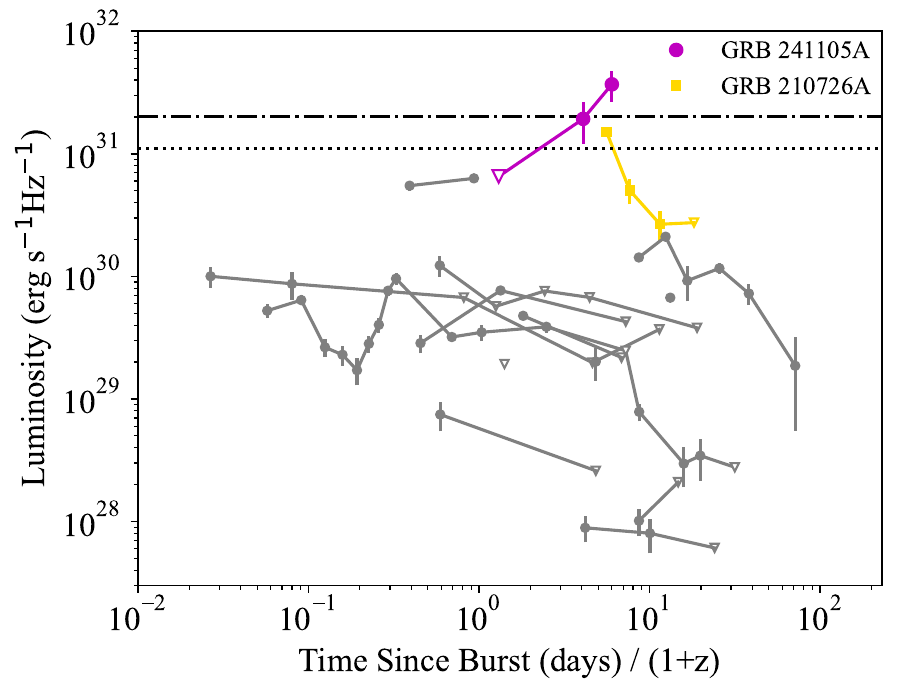}
    \caption{ Light curves of short GRBs in the X-ray ($0.3$–$10$~keV), optical ($R$-band), and radio ($8$–$10$~GHz) bands, shown in the \textbf{left}, \textbf{centre}, and \textbf{right} panels, respectively. In all panels, GRB241105A is highlighted in magenta, comparison short GRBs are shown in grey, and open triangles represent $3\sigma$ upper limits.
     \thisgrb exhibits the brightest optical afterglow among the short GRB population. 
     In the right panel, the dotted line shows the average long GRB luminosity of $1.1\times 10^{31}$\,erg\,s$^{-1}$\,Hz$^{-1}$ and dot dashed line shows the mean luminosity between $3-6$\,days post-burst in the rest frame of $\sim 2 \times 10^{31}$\,erg\,s$^{-1}$\,Hz$^{-1}$ \citep[][]{Chandra_2012}. 
    }
    \label{fig:Afterglow_comparison}
\end{figure*}
\subsection{A Bright Afterglow}\label{sec:afterglow_characteristics}
We compared the  X-ray (0.3 - 10 keV), optical ($R$-band), and radio light curves of \thisgrb with the short GRB population. The leftmost panel of Figure~\ref{fig:Afterglow_comparison} presents the X-ray light curves (in the $0.3$–$10$ keV band) for a sample of short GRBs, shown in grey, with GRB241105A overplotted in magenta. The flux light curves were taken from the \textit{Swift} XRT repository\footnote{\url{https://www.swift.ac.uk/xrt_curves/}} and converted to rest-frame luminosities using standard cosmological parameters. As is evident from the figure, GRB241105A exhibits significantly higher X-ray luminosity than the majority of short GRBs at comparable epochs. 

In the central panel of Figure~\ref{fig:Afterglow_comparison}, we show the rest-frame $R$-band light curves of short GRB afterglows, compiled from the datasets of \citet{fong15} and \citet{Rastinejad2021}, plotted in grey. GRB241105A is again shown in magenta. All magnitudes are corrected for Galactic extinction and converted to flux densities. Among the entire short GRB sample, GRB241105A stands out as the brightest known optical afterglow.

Only 18 confirmed short (merger driven) GRBs have been detected in the radio band \citep[][]{berger05,soderberg06,fong14,fong15,Fong21,Lamb19,Troja19,laskar22,chastain24,Levan24,Anderson_2024,Schroeder_2024,Schroeder_2025,Schroeder_2025GCN}, representing 13\% of the population \citep{Schroeder_2025}. Note that we include the long duration merger GRB 230307A \citep{Levan24} in this sample but not GW170817/GRB 170817A \citep{Abbott17_GRB}. 
Two-thirds of this radio-detected sample was detected within $1-2$\,days post-burst, most of which faded below detectability within 4-9\,days \citep{Anderson_2024}, indicating short GRBs usually have short-lived radio afterglows. Meanwhile, \thisgrb{} was undetected by ATCA at $\sim5$\,days, but then the rising radio afterglow was detected at $15$\,days (see Table~\ref{tab:atca}). GRB 210726A is the other example of a short GRB that was undetected in the radio band until $>10$\,days post-burst, at which point it underwent a radio flare between $11-62$\,days that was likely caused by energy injection or a reverse shock from a shell collision \citep{Schroeder_2024}. 
The evolution of the radio light curve of \thisgrb is quite different from that of the radio-detected short GRB population, which in nearly all cases are declining by $\sim6$ days in the rest frame when \thisgrb\ still appears to be brightening (see the left panel of Figure~\ref{fig:Afterglow_comparison}). 

\thisgrb\ is also more radio luminous than the radio-detected short GRB population.
As shown in the rightmost panel of Figure~\ref{fig:Afterglow_comparison}, the majority of these short GRBs have a radio luminosity $\lesssim10^{30}$\,erg\,s$^{-1}$\,Hz$^{-1}$, which is fainter than long GRBs \citep[radio luminosities usually between $\sim10^{30}-10^{32}$\,erg\,s$^{-1}$\,Hz$^{-1}$ up to $\sim100$\,days post-burst;][]{Chandra_2012,Anderson_2018}. 
Only \thisgrb\ and GRB 210726A are more luminous than the average radio luminosity of long GRBs of $1.1\times10^31$\,erg\,s$^{-1}$\,Hz$^{-1}$, with \thisgrb\ being the only one brighter than a canonical long GRB mean luminosity of $\sim2 \times 10^{31}$\,erg\,s$^{-1}$\,Hz$^{-1}$ between 3-6\,days in the rest frame \citep{Chandra_2012}. As a result, the radio luminosity and evolution of \thisgrb\ is more consistent with the long GRB radio-detected population.

The prompt efficiency ($\eta_{\rm \gamma}\approx75\%$; see Tables~\ref{tab:prompt_properties} and \ref{tab:afterglow-parameters}) of the whole burst is high but nevertheless consistent with the median $\eta_\gamma$ for both long- and short-duration GRBs \citep{zlp+07,fong15,lbm+15}. On the other hand, the isotropic-equivalent kinetic energy derived from multi-wavelength afterglow modelling is more typical of long-duration collapsar GRBs than of short, merger-driven bursts \citep{fong15}. Similarly, the circumburst density is orders of magnitude higher than the densities commonly inferred for compact-object merger environments \citep{fong15}. Such high circumburst densities are extremely rare for merger-driven GRBs, but are common in the actively star-forming, gas-rich environments of collapsar events \citep{Schulze_2011}.

\begin{figure*}
    \centering
    \includegraphics[width=\columnwidth]{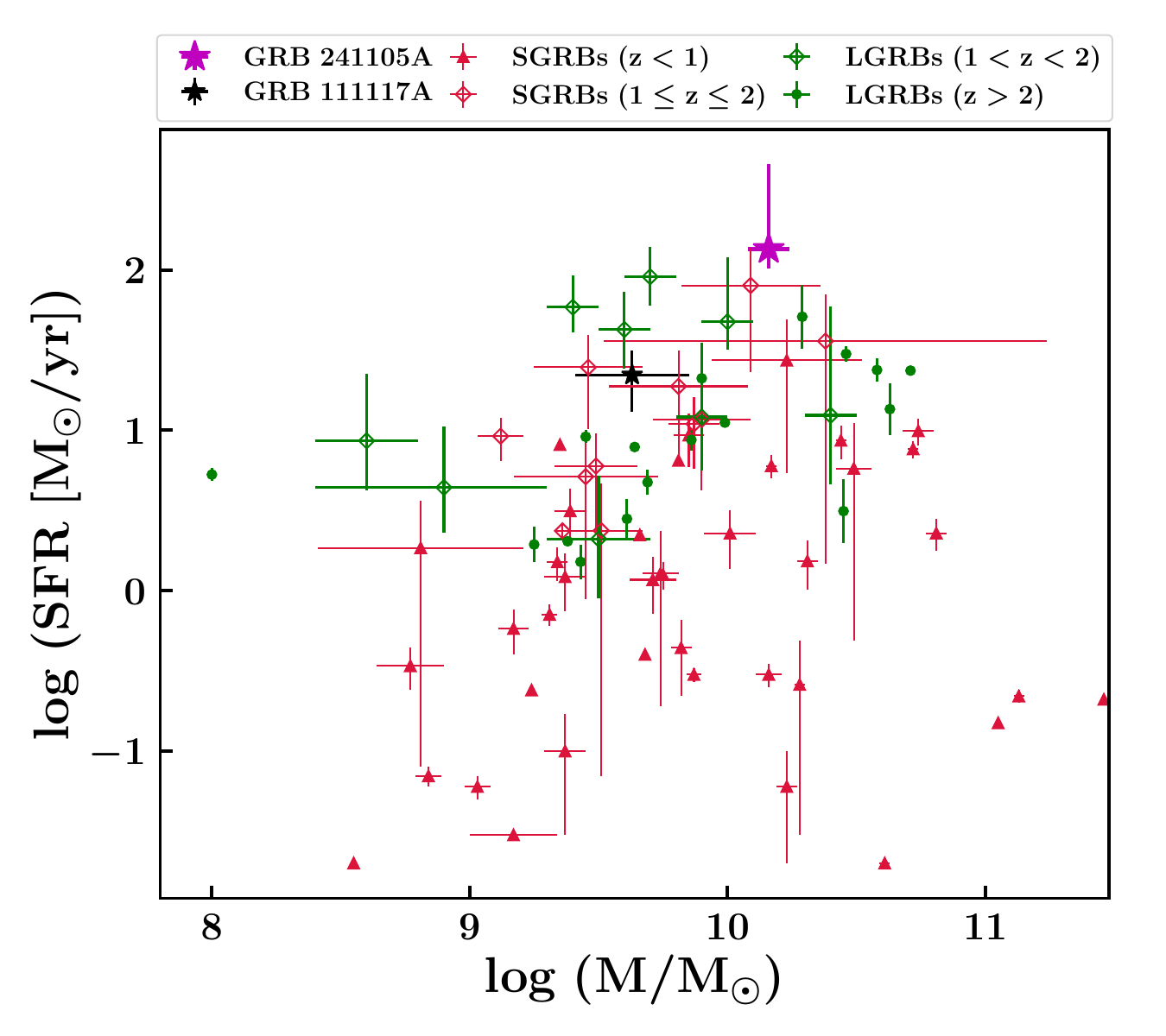}
    \includegraphics[width=\columnwidth]{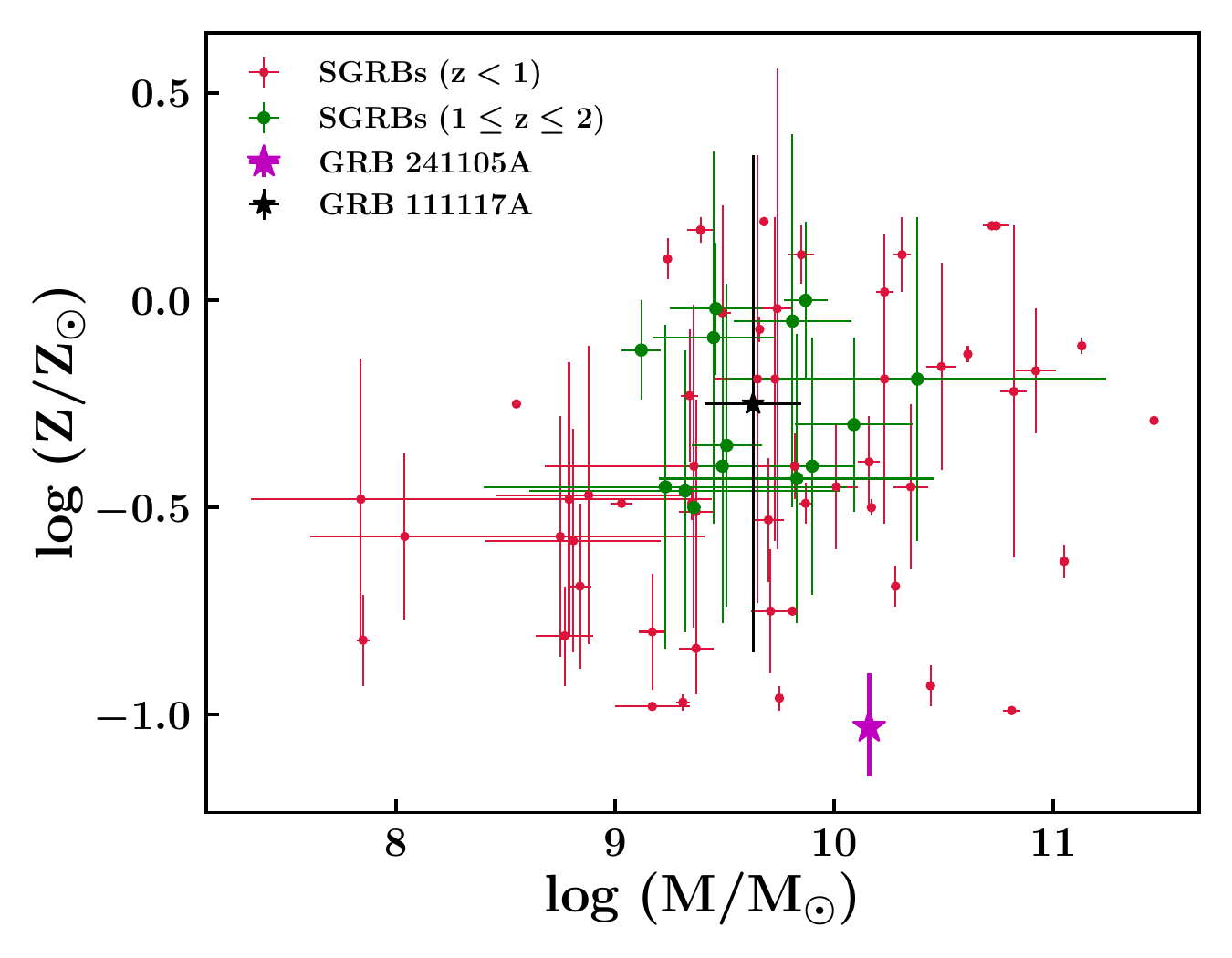}
    \caption{\textbf{\emph Left:} Mass versus SFR for \thisgrb (magenta star), compared with other short and long GRBs (Comparison data compiled from \citet{Nugent2024},\citet{Palmerio2019}, and \citet{Tanvir2019}. 
    \textbf{\emph Right:} M-Z relationship for \thisgrb as compared to other short GRBs compiled from \citet{Nugent2024}.}
    \label{fig:host_comp}
\end{figure*}
\subsection{A star-forming galaxy}
The host galaxy of \thisgrb is moderately massive, actively star-forming, and metal-poor. We compared the host properties of \thisgrb with the short GRB population using data from \cite{Nugent2024}. The host has quite high star formation rate as compared to short GRB hosts (see left panel of Figure~\ref{fig:host_comp}). We further included long GRB hosts from \cite{Palmerio2019} and \cite{Tanvir2019}. At this stellar mass and redshift, only about 15\% of star forming galaxies have similarly high star formation rates \citep{Whitaker12}, indicating rapidly forming stellar populations. Further, comparison of metallicity with the short GRBs reveals that the host has relatively low metallicity as compared to the short GRBs (see the right panel of Figure~\ref{fig:host_comp}). The metallicity is lower ($\ sim$0.5 dex) than expected from the mass-star formation rate-metallicity relation \citep{Mannucci10, Dave12}, consistent with the overall young age. The burst has a small offset from the host center, and the host's irregular, clumpy morphology and structural parameters are typical of collapsar hosts. The observed offset of \thisgrb aligns more closely with expectations for collapsar progenitors than for compact object mergers \citep[which generally occur at larger distances due to natal kicks;][] {Narayan_1992, Bloom_1999, Perna_2002}.

We note that another short GRB~090426 ($T_{90} = 1.28$ s, $z = 2.609$), exhibited a host galaxy that was blue, luminous, and actively star-forming with a small angular offset—characteristics more typical of long GRBs from collapsar progenitors \citep{Antonelli_2009}, and similar to what we observe for GRB 241105A. However, this is not the case for all high-redshift short bursts; for example, GRB 111117A ($z = 2.211$) had host galaxy properties and an offset from its host center that are more consistent with merger-driven GRBs \citep{Selsing_2018}. These contrasting examples highlight that, while host galaxy properties can offer valuable clues, they alone are not always definitive for distinguishing between progenitor scenarios, especially at high redshift.

\section{Summary}
We present a comprehensive multi-wavelength analysis of \thisgrb, an ambiguous GRB, exhibiting both short and long GRB characteristics. The key results of our study are summarised below:
\begin{itemize}
\item The prolonged emission episode of GRB 241105A is spectrally harder than the short, initial pulse. This contrasts with the typical behaviour of extended-emission GRBs from mergers, where the later emission is generally softer than the initial spike.
\item Initial emission pulse exhibits a short minimum variability timescale, aligning with expectations for compact central engines in short GRBs, while the comparatively longer variability timescale of the later emission episode points to extended activity.
\item The spectral lag of the initial short pulse is consistent with the lag-luminosity relation, which is usually seen in the case of long GRBs. 
\item In the Amati plane, the initial short pulse occupies a position near the overlapping region between short and long GRBs. In contrast, the prolonged emission and the whole burst are well located within the short GRB plane.
\item PCA-UMAP clustering places \thisgrb near both long collapsars and long-duration mergers, suggesting shared temporal and spectral characteristics with both populations. This proximity suggests shared characteristics, possibly in central engine behavior (e.g., black hole accretion or magnetar activity) or emission mechanisms (e.g., prolonged jet activity or synchrotron radiation), rather than a definitive progenitor type. 
\item The afterglow emission is exceptionally luminous in the X-ray, optical, and radio bands compared to the majority of short GRBs. Modelling requires a high-density circumburst medium, implying a dense, star-forming environment consistent with collapsar progenitors. 
\item The host galaxy is moderately massive, actively star-forming, and metal-poor, with a star formation rate higher than that observed in typical short GRB hosts. The burst exhibits a small projected offset, which also favours collapsar.
\end{itemize}
The prompt emission characteristics of \thisgrb present a complex and nontrivial picture. While the host galaxy properties and the burst environment lend support to a collapsar origin, they do not definitively exclude a compact-binary merger scenario. Notably, the presence of a star-forming environment is not in itself conclusive: GRB~060505, for example, occurred in a vigorously star-forming galaxy, coincided with an HII region showed a spectral lag, yet no supernova was detected \citep{Fynbo06, McBreen08, Xu_2009}. Such cases suggest the existence of supernova-less collapsars or alternative progenitor channels and caution against relying solely on host galaxy properties for classification. 

A compact-binary merger origin remains possible; population simulations of neutron star mergers indicate that merger transients at $z = 2.681$ would have an event probability of $\sim 1$-in-$1000$ for \swift detected short GRBs, and would likely be well associated with its host galaxy \citep[i.e., have a low impact parameter;][]{Mandhai2022}. Thus, both progenitor scenarios remain viable for \thisgrb. If this GRB does indeed have a compact-binary merger origin, it would provide observational support for models invoking rapid, early-universe binary evolution and short delay times, which would help to investigate the chemical evolution of the early Universe. 
On the other hand, if \thisgrb originated from a collapsar, it may resemble a growing set of high-redshift events that exhibit a bright initial short spike followed by weaker extended emission. Several such GRBs, identified by \citet{Dichiara_2021, Dimple2022, Dimple2024}, occur in collapsar-like environments, yet mimic the short+EE light-curve morphology. However, in these cases, the extended emission is detected predominantly in the higher-energy bands. \thisgrb may therefore belong to this ambiguous subset of bursts, suggesting that some similar events might have been misclassified and highlighting the need for a careful re-examination of GRB classification criteria.

\thisgrb thus sits at the boundary of traditional GRB frameworks, emphasizing the need for more nuanced classification schemes. Whether it marks the most distant compact-binary merger or a rare collapsar with atypical properties, its ambiguous nature makes it a benchmark case for progenitor identification and for probing star formation and chemical enrichment in the early Universe.

\section*{Acknowledgments}
Dimple and BPG acknowledge support from STFC grant No. ST/Y002253/1. BPG and DO acknowledge support from the Leverhulme Trust grant No. RPG-2024-117. Dimple acknowledges IUCAA, India, for their warm hospitality during her 
stay on the campus. The Gravitational-wave Optical Transient Observer (GOTO) project acknowledges support from the Science and Technology Facilities Council (STFC, grant numbers ST/T007184/1, ST/T003103/1, ST/T000406/1, ST/X001121/1 and ST/Z000165/1) and the GOTO consortium institutions; University of Warwick; Monash University; University of Sheffield; University of Leicester; Armagh Observatory \& Planetarium; the National Astronomical Research Institute of Thailand (NARIT); University of Manchester; Instituto de Astrofísica de Canarias (IAC); University of Portsmouth; University of Turku. AAC acknowledges support through the European Space Agency (ESA) research fellowship programme. AS acknowledges support by a postdoctoral fellowship from the CNES. A Kumar is supported by the UK Science and Technology Facilities Council (STFC) Consolidated grant ST/V000853/1. T.E.M.B. is funded by Horizon Europe ERC grant no. 101125877. KT acknowledges support from the Dynamical Universe, funded by the KAW foundation. S.M. acknowledges financial support from the Research Council of Finland project 350458. A. Sahu acknowledges the Warwick Astrophysics PhD prize scholarship, made possible thanks to a generous philanthropic donation. JDL acknowledges support from a UK Research and Innovation Future Leaders Fellowship (MR/T020784/1). R.B. acknowledges funding from the Italian Space Agency, contract ASI/INAF n. I/004/11/6. S.B., O.M., and O.R. gratefully acknowledge NASA funding from cooperative agreement 80NSSC24M0035. CdB and SMB acknowledge support from funded by Taighde Éireann -- Research Ireland under Grant numbers 21/FFP-A/9043. PV gratefully acknowledges NASA funding from cooperative agreement 80MSFC22M0004. RLCS acknowledges support from The Leverhulme Trust grant No. RPG-2023-240. A.R. acknowledges support from INAF minigrant 1.05.24.07.04. RB, SC, PDA, AM and RS acknowledge funding from the Italian Space Agency, contract ASI/INAF n. I/004/11/6.  M.E.R. is supported by the European Union (ERC, Starstruck, 101095973, PI Jonker). TLK acknowledges a Warwick Astrophysics prize postdoctoral fellowship made possible thanks to a generous philanthropic donation. IW and MEW are supported by the UKRI Science and Technology Facilities Council (STFC). This work is based [in part] on observations made with the NASA/ESA/CSA James Webb Space Telescope. The data were obtained from the Mikulski Archive for Space Telescopes at the Space Telescope Science Institute, which is operated by the Association of Universities for Research in Astronomy, Inc., under NASA contract NAS 5-03127 for JWST. These observations are associated with program 9228. We thank the referee for their careful reading of the manuscript and for constructive comments that helped to improve the clarity and quality of this work.

\section*{Data Availability}
The gamma-ray, X-ray, optical, infrared, and radio datasets underlying this article are based on observations collected from public archives and instrument teams, as described in Section~\ref{sec:observations}. \gbm, \bat, \xrt, and \uvot data are publicly available from the HEASARC archive.\footnote{\url{https://heasarc.gsfc.nasa.gov/}} \grm data will be made available upon reasonable request from the instrument team. JWST data are accessible from the Mikulski Archive for Space Telescopes (MAST)\footnote{\url{https://mast.stsci.edu/}} under program ID 9228. Ground-based optical and radio data (including GOTO, VLT/FORS2, NTT, and ATCA) can be obtained from the respective observatories or are available from the corresponding author upon reasonable request.



\bibliographystyle{mnras}
\bibliography{main,agmodeling} 


\newpage
\section*{Affiliations}
$^{1}$School of Physics and Astronomy, University of Birmingham, Edgbaston, Birmingham, B15 2TT, UK.\\
$^{2}$Institute for Gravitational Wave Astronomy, University of Birmingham, Birmingham, B15 2TT, UK.\\
$^{3}$Department of Astrophysics/IMAPP, Radboud University, 6525 AJ Nijmegen, The Netherlands.\\
$^{4}$Science and Technology Institute, Universities Space Research Association, Huntsville, AL 35805, USA.\\
$^{5}$European Space Agency (ESA), European Space Research and Technology Centre (ESTEC), Keplerlaan 1, 2201 AZ Noordwijk, The Netherlands.\\
$^{6}$Niels Bohr Institute, University of Copenhagen, Jagtvej 128, 2200 Copenhagen N, Denmark.\\
$^{7}$INAF, Osservatorio Astronomico di Capodimonte, Salita Moiariello 16, I-80131, Naples, Italy.\\
$^{8}$Astrophysics Research Institute, Liverpool John Moores University, IC2 Liverpool Science Park, 146 Brownlow Hill, Liverpool, L3 5RF, UK.\\
$^{9}$Department of Physics \& Astronomy, University of Utah, Salt Lake City, UT 84112, USA.\\
$^{10}$Université Paris-Saclay, Université Paris Cité, CEA, CNRS, AIM, 91191, Gif-sur-Yvette, France.\\
$^{11}$International Centre for Radio Astronomy Research, Curtin University, GPO Box U1987, Perth, WA 6845, Australia.\\
$^{12}$School of Physics, Centre for Space Research, Science Center North, University College Dublin, Dublin 4, Ireland.\\
$^{13}$State Key Laboratory of Particle Astrophysics, Institute of High Energy Physics, Chinese Academy of Sciences, Beijing 100049, China.\\
$^{14}$Department of Physics, Royal Holloway - University of London, Egham Hill, Egham, TW20 0EX, UK.\\
$^{15}$Key Laboratory of Space Astronomy and Technology, National Astronomical Observatories, Chinese Academy of Sciences, Beijing 100101, China.\\
$^{16}$Department of Physics, Lancaster University, Lancaster, LA1 4YB, UK.\\
$^{17}$Department of Physics, Indian Institute of Technology Bombay, Powai, Mumbai 400076, India.\\
$^{18}$School of Physics, Indian Institute of Science Education and Research Thiruvananthapuram, Thiruvananthapuram, 695551, India.\\
$^{19}$ Department of Space Science and Center for Space Plasma and Aeronomic Research, University of Alabama in Huntsville, Huntsville, AL 35899, USA.\\
$^{20}$Department of Physics, University of Warwick, Coventry CV4 7AL, UK.\\
$^{21}$INAF, Osservatorio Astronomico di Brera, Via E. Bianchi 46, 23807 Merate (LC), Italy.\\
$^{22}$INAF--Osservatorio di Astrofisica e Scienza dello Spazio, via Piero Gobetti 93/3, 40024, Bologna, Italy.\\
$^{23}$School of Physics and Astronomy, University of Leicester, University Road, LE1 7RH Leicester, United Kingdom.\\
$^{24}$Excellence Cluster ORIGINS, Boltzmannstra\ss e 2, 85748 Garching, Germany.\\
$^{25}$Ludwig-Maximilians-Universit\"at, Schellingstra\ss e 4, 80799 M\"unchen, Germany.\\
$^{26}$Jodrell Bank Centre for Astrophysics, University of Manchester, Manchester M13 9PL, UK.\\
$^{27}$Instituto de Astrof\'isica de Canarias, E-38205 La Laguna, Tenerife, Spain.\\
$^{28}$Departamento de Astrof\'isica, Universidad de La Laguna, E-38206 La Laguna, Tenerife, Spain.\\
$^{29}$Department of Physics and Astronomy, University of New Mexico, 210 Yale Blvd NE, Albuquerque, NM 87106, USA.\\
$^{30}$Space Science Data Center (SSDC) - Agenzia Spaziale Italiana (ASI), I-00133 Roma, Italy.\\
$^{31}$School of Mathematical and Physical Sciences, University
of Sheffield, Sheffield S3 7RH, UK.\\
$^{32}$School of Physics and Astronomy, Monash University, Clayton, VIC 3800, Australia.\\
$^{33}$Sydney Institute for Astronomy, School of Physics, The University of Sydney, NSW 2006, Australia.\\
$^{34}$ARC Centre of Excellence for Gravitational Wave Discovery (OzGrav), Hawthorn, VIC 3122, Australia.\\
$^{35}$CSIRO Space and Astronomy, PO Box 76, Epping, NSW 1710, Australia.\\
$^{36}$Department of Physics and Astronomy, Clemson University, Clemson, SC 29634, USA.\\
$^{37}$Center for Astrophysics and Cosmology, Science Institute, University of Iceland, Dunhagi 5, 107 Reykjavik, Iceland.\\
$^{38}$Department of Physics and Astronomy, University of Turku, FI-20014 Turku, Finland.\\
$^{39}$David A. Dunlap Department of Astronomy and Astrophysics, University of Toronto, 50 St. George Street, Toronto, ON M5S 3H4, Canada.\\
$^{40}$Dunlap Institute for Astronomy and Astrophysics, University of Toronto, 50 St. George Street, Toronto, ON M5S 3H4, Canada.\\
$^{41}$Racah Institute of Physics, The Hebrew University of Jerusalem, Jerusalem 91904, Israel.\\
$^{42}$INAF, Osservatorio Astronomico di Roma, via Frascati 33, I-00078 Monte Porzio Catone (Roma), Italy.\\
$^{43}$School of Sciences, European University Cyprus, Diogenes Street, Engomi, 1516 Nicosia, Cyprus.\\
$^{44}$School of Mathematical and Physical Sciences, Macquarie University, NSW 2109, Australia.\\
$^{45}$Astrophysics and Space Technologies Research Centre, Macquarie University, Sydney, NSW 2109, Australia.\\
$^{46}$National Astronomical Research Institute of Thailand (NARIT), Chiang Mai 50180, Thailand.\\
$^{47}$Institute of Cosmology and Gravitation, University of Portsmouth, Portsmouth PO1 3FX, UK.\\
$^{48}$Department of Mathematics and Computer Sciences, Physical Sciences and Earth Sciences, University of Messina, F.S. D'Alcontres 31, 98166, Messina, Italy.\\
$^{49}$Anton Pannekoek Institute of Astronomy, University of Amsterdam, Science Park 904, 1098 XH Amsterdam, The Netherlands.\\
$^{50}$Armagh Observatory \& Planetarium, College Hill, Armagh BT61 9DB, UK.\\
$^{51}$INAF---Istituto di Astrofisica Spaziale e Fisica Cosmica di Milano, Via A. Corti 12, 20133 Milano, Italy\\
$^{52}$Department of Physics, University of Bath, Claverton Down, Bath, BA2 7AY, UK.\\
$^{53}$Aix Marseille University, CNRS, CNES, LAM, Marseille, France.\\
$^{54}$Department of Physics, George Washington University, 725 21st St NW, Washington, DC 20052, USA.\\
$^{55}$School of Physics, Trinity College Dublin, The University of Dublin, Dublin 2, Ireland.\\
$^{56}$Instituto de Ciencias Exactas y Naturales (ICEN), Universidad Arturo Prat, Chile.\\
$^{57}$ The Oskar Klein Centre, Department of Astronomy, Stockholm University, AlbaNova, SE-10691, Stockholm, Sweden.\\
$^{58}$ European Southern Observatory, Alonso de Córdova 3107, Casilla 19, Santiago, Chile.\\
$^{59}$ Instituto de Alta Investigaci\'on, Universidad de Tarapac\'a, Casilla 7D, Arica, Chile.\\
$^{60}$ Astronomical Observatory, University of Warsaw, Al. Ujazdowskie 4, 00-478 Warszawa, Poland.\\
$^{61}$ Centre for Astrophysics Research, University of Hertfordshire, College Lane, Hatfield, AL10 9AB, United Kingdom.\\
$^{62}$ Department of Physics \& Astronomy, Louisiana State University, Baton Rouge, LA 70803, USA.\\
$^{63}$ Millennium Institute of Astrophysics MAS, Nuncio Monsenor Sotero Sanz 100, Off. 104, Providencia, Santiago, Chile.\\
$^{64}$University of Chinese Academy of Sciences, Chinese Academy of Sciences, Beijing 100049, China.\\
$^{65}$Centre for High Performance Computing, Indian Institute of Science Education and Research Thiruvananthapuram, Thiruvananthapuram, 695551, India.\\

\newpage
\appendix
\section{Wind model}
\label{appendix:wind}
\begin{figure*}
    \centering
    \begin{tabular}{cc}
         \includegraphics[width=\columnwidth]{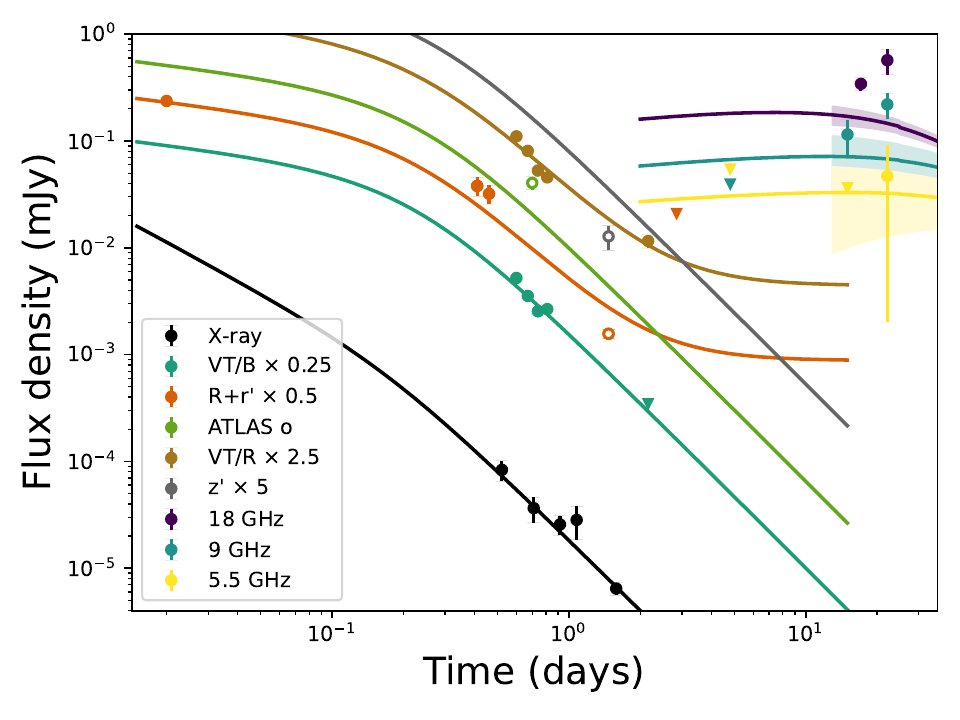} &  \includegraphics[width=\columnwidth]{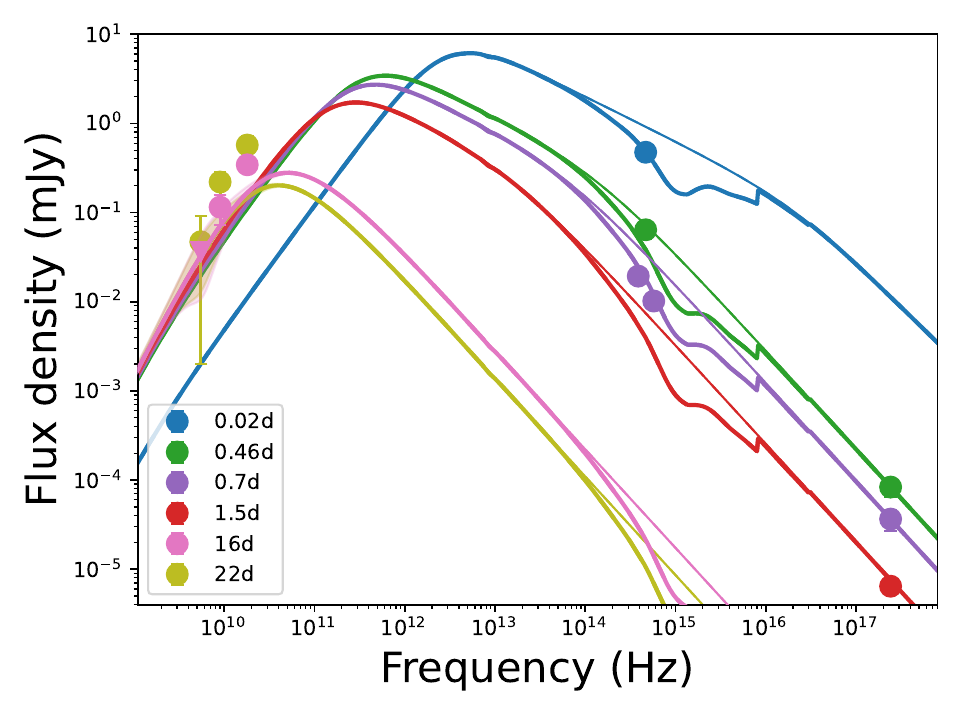}
    \end{tabular}    
    \caption{Light curves (left) and spectral energy distributions (SEDs, right) for our best-fit wind model for the afterglow observations of GRB~241105A. Data points plotted in open symbols are not included in the analysis. Shaded bands indicate $1\sigma$ variability at each time expected from interstellar scintillation for reference. The flattening in the VT/R-band and $R+r^\prime$-band light curves is due to a fixed host contribution of $1.8\,\mu$Jy included in the modeling. Correlation contours for all physical parameters included in the fit, along with the derived parameters of  $\theta_{\rm jet}$ and $E_K$, are presented in Figure~\ref{fig:ag-corner-wind} and the corresponding model is discussed in Appendix~\ref{appendix:wind}. See Figure~\ref{fig:ag-ISM} for the corresponding light curve and SED plots for the ISM model.}
    \label{fig:ag-wind}
\end{figure*}

\begin{figure*}
    \centering
    \includegraphics[width=0.9\textwidth]{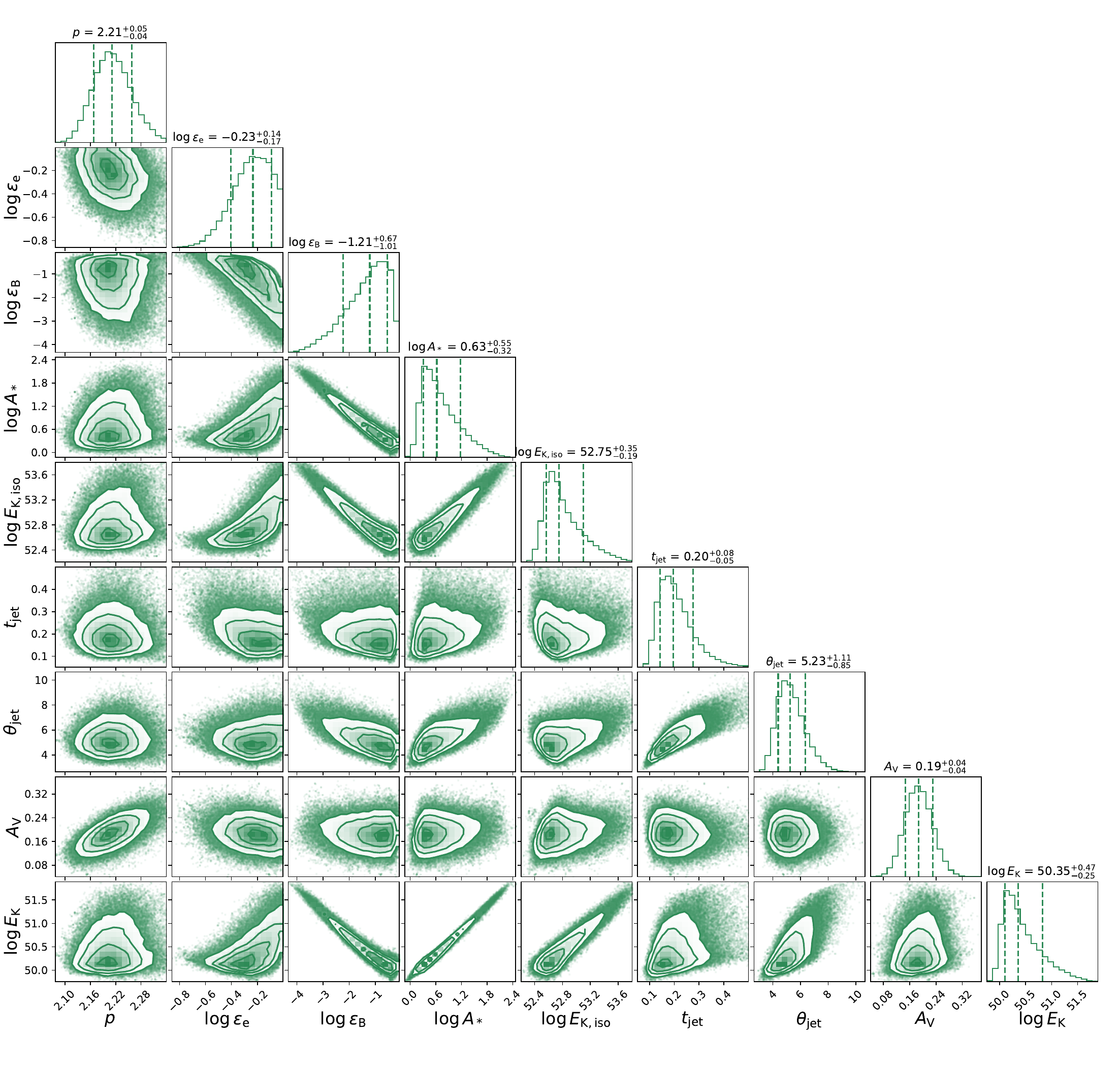}
    \vspace{-0.3in}
    \caption{Correlations and marginalised posterior density for all free parameters in the afterglow model in a wind environment. $E_{\rm K,iso}$ and $E_K$ are in units of erg, $t_{\rm jet}$ in days, $\theta_{\rm jet}$ in degrees, and $A_V$ in magnitudes. The contours enclose 39.3\%, 86.4\%, and 98.9\% of the probability mass in each correlation plot (corresponding to 1, 2, and 3$\sigma$ regions for two-dimensional Gaussian distributions, respectively. The dashed lines indicate the 15.9\%, 50\%, and 84.1\% quantiles, corresponding to the median and $\pm1\sigma$ for one-dimensional Gaussian distributions. See Figure~\ref{fig:ag-corner-ISM} for the corresponding corner plot for the ISM model.}
    \label{fig:ag-corner-wind}
\end{figure*}

As discussed in Section~\ref{text:prelim-optx}, the observations are degenerate with respect to the circumstellar density profile, and thus both ISM and wind models are feasible. The ISM model is discussed in Section~\ref{text:mcmc}. In this section, we discuss the wind model. We perform an MCMC analysis identical to that described in Section~\ref{text:mcmc}. Our best-fit wind model has $p=2.21$, $\epsilon_{\rm e} = 0.49$, $\epsilon_{\rm B} = 0.15$, $A_* = 2.5$, $E_{\rm K,iso}=4.6\times10^{52}$~erg, $t_{\rm jet}=0.17$~days, and $A_V=0.20$~mag. 
For these parameters, $\nu_{\rm m}\approx2.4\times10^{15}$~Hz at $\approx0.1$~days, passing through the optical at 0.1--0.4~days, similar to the ISM case. The afterglow is again in fast cooling with $\nu_{\rm c}<\nu_{\rm m}$. The remaining break frequencies for the best-fit wind model are listed in Table~\ref{tab:afterglow-parameters}. 
We plot light curves and SEDs from the radio to X-rays of the best-fit model in Figure~\ref{fig:ag-wind} and provide corner plots of the correlations and marginalised posterior density for all fitted parameters (and the derived parameters, $\theta_{\rm jet}$ and $E_K$) from our MCMC analysis in Figure~\ref{fig:ag-corner-wind}. We note that the wind model fit results in a slightly poorer $\chi^2_{\rm wind}=43$ (18 degrees of freedom) compared with that for the ISM model ($\chi^2_{\rm ISM}=41$, 18 d.o.f) due to a marginally poorer fit to the radio data, but the two models are otherwise indistinguishable. We also note that both models result in very similar values for most of the model parameters. The sole exception is $\epsilon_B$, which is, however, strongly degenerate with both the energy and the density (both of which allow for acceptable fits over a range of at least one order of magnitude) due to the unconstrained cooling break. Despite this, the beaming-corrected kinetic energy is consistent at $\log E_{\rm K}\approx50.3$ between both cases. 

\section{Curve of Growth}
\label{sec:CoG}
\begin{figure}
    \centering
    \includegraphics[scale=0.6]{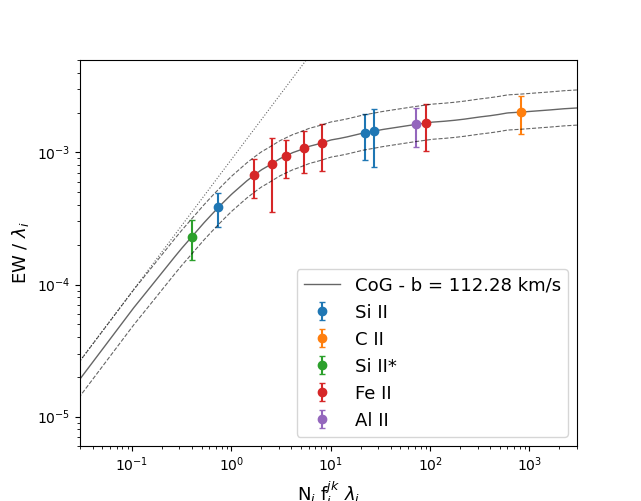}
    \caption{The results of the curve of growth analysis. Absorption lines from the same ionized state have been reported with the same color. Confidence limits, dashed lines, have been provided at 2 sigma level. The dotted line corresponds to the linear case, where lines are not considered saturated.}
    \label{fig:CoG}
\end{figure}

We conducted the CoG analysis following \citet{Spitzer1998}.  We identified distinct absorption lines of low-excitation electronic transitions, unblended with other absorption lines or telluric features, to measure the equivalent width (EW) of the absorption lines. The \ion{Si}{II} 1260,1808 \AA\, lines from the ground state were then used to establish the Doppler parameter, resulting in $b = 112.3 \pm 35.8$\,km\,s$^{-1}$, which aligns well with the Voigt fit analysis. A comprehensive explanation of this methodology can be found in the examination of the high-redshift ultra-long GRB 220627A \citep{deWet2023}. 
Subsequently, we conducted a fit incorporating the entire set of absorption lines listed in Table~\ref{tab:CoG}, which exhibit very similar excitation energies, under the assumption that the particles responsible for the absorption features are located in ISM clouds with similar velocity distributions (same $b$ parameter). Although higher-excitation features were detected in the spectrum of \thisgrb, only \ion{Si}{IV} 1393,1404 \AA\, were not blended with other features, so we skip the CoG analysis for these high-excitation lines. The findings of our analysis are presented in Table~\ref{tab:CoG} and Figure \ref{fig:CoG}. Moreover, it is worth noting that the relatively high column density of \ion{C}{II} 1334 \AA\, is likely influenced by the presence of an underlying \ion{C}{II*} transition, which enhances the EW measured for this line and the corresponding column density. 
It is important to remark that the majority of the absorption features in this analysis fall in the flat/saturated regime of the CoG, indicating that the column densities obtained through this methodology should be approached with caution and considered as lower limits to the actual density of absorbers in the ISM clouds \citep{Prochaska2006}. The low spectral resolution of FORS2 primarily drives this conclusion. At an average resolving power of $R = 440$ (grism 300V) and $R = 660$ (grism 300I), the dispersion velocity resolution measured is approximately $\delta v \approx 550$\,km\,s$^{-1}$ at 6655 \AA\,, which corresponds to the observed wavelength of \ion{Si}{II} 1808 \AA\,. This implies that we cannot resolve individual ISM clouds, particularly those located in the immediate vicinity of the GRB. Therefore, each absorption line in this analysis involves contributions from multiple, if not all, clouds within the GRB host galaxy, reinforcing the notion that our measurements should be considered lower limits.
Furthermore, we did not perform a more detailed metallicity and dust depletion analysis of the ISM given these limitations.

\begin{table}
  \centering
  \caption{Column densities derived via the curve of growth method for various ionic transitions. Equivalent widths (EW) and observed wavelengths are reported alongside their respective column densities, with uncertainties reflecting 1$\sigma$ confidence intervals.}
  \renewcommand{\arraystretch}{1.4}
  \begin{threeparttable}
    \begin{tabular}{l c c c}
      \hline
      Ion & $\lambda_{\rm obs}$ (\AA) & EW$_{\lambda}$ (\AA) & $\log N_X$ \\
      \hline
      \ion{Si}{II} 1260   & 4639.6 & $6.74 \pm 3.17$ & $15.90^{+0.53}_{-0.53}$ \\
      \ion{Si}{II} 1527   & 5619.8 & $7.93 \pm 3.01$ & $15.90^{+0.53}_{-0.53}$ \\
      \ion{Si}{II} 1808   & 6655.3 & $2.55 \pm 0.73$ & $15.90^{+0.53}_{-0.53}$ \\
      \ion{C}{II} 1334    & 4912.4 & $9.93 \pm 3.18$ & $16.84^{+0.80}_{-1.00}$ \\
      \ion{Si}{II*} 1533  & 5644.6 & $1.30 \pm 0.44$ & $14.10^{+0.18}_{-0.27}$ \\
      \ion{Fe}{II} 1608   & 5920.7 & $3.99 \pm 1.31$ & $15.43^{+0.32}_{-0.31}$ \\
      \ion{Fe}{II} 2344   & 8629.0 & $10.22 \pm 3.95$ & $15.43^{+0.32}_{-0.31}$ \\
      \ion{Fe}{II} 2374   & 8740.4 & $8.22 \pm 2.65$ & $15.43^{+0.32}_{-0.31}$ \\
      \ion{Fe}{II} 2383   & 8770.9 & $7.23 \pm 4.12$ & $15.43^{+0.32}_{-0.31}$ \\
      \ion{Fe}{II} 2587   & 9521.4 & $10.21 \pm 3.60$ & $15.43^{+0.32}_{-0.31}$ \\
      \ion{Fe}{II} 2600   & 9571.2 & $15.98 \pm 6.25$ & $15.43^{+0.32}_{-0.31}$ \\
      \ion{Al}{II} 1671   & 6150.2 & $10.01 \pm 3.22$ & $15.55^{+1.26}_{-1.07}$ \\
      \hline
    \end{tabular}
  \end{threeparttable}
  \label{tab:CoG}
\end{table}


\label{lastpage}
\end{document}